\documentclass[12pt,prd,aps,amssymb,amsmath,tightenlines,showpacs]{article}
\usepackage[utf8]{inputenc}
\usepackage[a4paper,top=3cm,bottom=3cm,left=2.5cm,right=2.5cm]{geometry}
\usepackage{amssymb}
\usepackage{amsmath}
\usepackage{amsfonts}
\usepackage{mathtools}
\usepackage{amsthm}
\usepackage{tipa}
\usepackage{yhmath} 
\usepackage{latexsym} 
\usepackage{graphicx}
\usepackage{comment}
\usepackage{color}
\usepackage{tikz}
\usepackage{cite}
\usepackage[compat=1.1.0]{tikz-feynman}

\newcommand{\be}{\begin{equation}}
\newcommand{\ee}{\end{equation}}
\newcommand{\bea}{\begin{eqnarray}}
\newcommand{\eea}{\end{eqnarray}}

\begin{document}
\graphicspath{{FIGURE/}}
\topmargin=0.0cm

\begin{center} %{\large{\bf Original study of the  $g \phi^2 (i\phi)^{\epsilon}$ theory\\Analysis at all orders in $\epsilon$ and resummations}}\\	
{\large{\bf Original study of the   $g \phi^2 (i\phi)^{\epsilon}$ theory\\ 
Analysis of all orders in $\epsilon$ and resummations}}\\
%{\large{\bf Logarithmic expansion of the $g \phi^2 (i\phi)^{\epsilon}$ theory\\Analysis at all orders in $\epsilon$ and resummations}}

	\vspace*{0.8 cm}
		
	Vincenzo Branchina$^{a,b,}$\,\footnote{branchina@ct.infn.it}, 
	Alberto Chiavetta$^{a,c,}$\,\footnote{albertochiavetta@gmail.com}, 
	Filippo Contino$^{a,b,}$\,\footnote{filippo.contino@ct.infn.it}
	\vspace*{0.4cm}
	
	{\it ${}^a$Department of Physics and Astronomy, University of Catania,\\
		Via Santa Sofia 64, 95123 Catania, Italy,} \\
	\vspace*{0.02cm}
	{\it ${}^b$
		INFN, Sezione di Catania, Via Santa Sofia 64, 95123 Catania, Italy,} \\
	\vspace*{0.02cm}
	{\it ${}^c$Scuola Superiore di Catania, Via Valdisavoia 9, 95123 Catania, Italy} \\
	
	\vspace*{1 cm}

	{\LARGE Abstract}\\
\end{center}

In a recent work the Green's functions of the $\mathcal{PT}$-symmetric scalar theory $g \phi^{2}(i\phi)^\epsilon$ were calculated at the first order  of the logarithmic expansion, i.e.\,at first order in $\epsilon$, and it was proposed to use this expansion in powers of $\epsilon$ to implement a systematic renormalization of the theory. Using techniques that we recently developed for the analysis of an ordinary (hermitian) scalar theory,  in the present work we calculate the Green's functions at $\mathcal{O}(\epsilon^2)$, pushing also the analysis to higher orders. We find that, {\it at each finite order in $\epsilon$, the theory is non-interacting for any dimension $d \geq 2$}.
We then conclude that by no means
this expansion can be used for a systematic renormalization of the theory. We are then lead to consider resummations, and we start with the leading contributions.
Unfortunately, the results are quite poor. Specifying to the physically relevant $i g \phi^3$ model, we show that this resummation simply gives the trivial lowest order results of the weak-coupling expansion. We successively resum subleading diagrams, but again the results are rather poor. All this casts serious doubts on the possibility of studying the theory $g \phi^{2}(i\phi)^\epsilon$ with the help of such an expansion. 
We finally add that the findings  presented in this work were obtained by us some time ago (December 2019),
and we are delighted to see that these results, that we communicated to C.M. Bender in December 2019, are confirmed in a recent preprint (e-Print:2103.07577) of C.M. Bender and collaborators.   

\section{Introduction}

The $\mathcal{PT}$-symmetric scalar theory $g \phi^2(i \phi)^\epsilon$ was recently studied within the framework of the 
logarithmic expansion, that is an expansion in powers of $\epsilon$.
%, previously used for the ordinary scalar theory $g\phi^2(\phi^2)^{\epsilon}$\cite{Bender:1987dn,Bender:1988rq}. 
The Green's functions $\mathcal G_n$ were calculated at first order in $\epsilon$, and it was proposed to use this expansion to implement a systematic renormalization of the theory\cite{Bender:2018pbv}.

\begin{comment}

A logarithmic expansion for 
the scalar theory
$g\phi^2(\phi^2)^{\epsilon}$,   i.e. an expansion in terms of the parameter  $\epsilon$, was proposed in\,\cite{Bender:1987dn,Bender:1988rq}, and the Green's functions $\mathcal G_n$ were calculated at first order in $\epsilon$.
This expansion was recently reconsidered\cite{Bender:2018pbv} for studying the $\mathcal{PT}$-symmetric scalar theory $g \phi^2(i \phi)^\epsilon$; the
$\mathcal G_n$  were calculated at first order in $\epsilon$, and it was proposed to use it to implement a systematic renormalization of the theory.
\end{comment}

Of great physical interest is the case $\epsilon=1$, the $ig\phi^3$ model,  introduced some time ago by Fisher\cite{Fisher:1978pf} to study the density of the Lee-Yang zeros of the partition function $Z[h]$ on  the imaginary axis of $h$ ($h$ is the magnetic field). Its renormalization properties were investigated in\cite{Bender:2012ea,Bender:2013qp}, where the critical exponents were calculated. This model 
is the field theoretic analogue of the $\mathcal{PT}$-symmetric quantum-mechanical theory described by the
Hamiltonian $H = p^2 + ix^3$, whose eigenvalues have been rigorously shown to be all
real\cite{Dorey:2001uw}.
%and for this reason it is highly regarded when considering the extension of PT-symmetry to field theories
While in the case of $\mathcal{PT}$-symmetric quantum mechanics, however, the boundary conditions on the Schrodinger equation are imposed in complex Stokes sectors\cite{Bender:2007nj}, the search for the Stokes wedges for the integration over infinitely many complex field variables is an insurmountable task. 
The advancement that the expansion in $\epsilon$ seems to bring %in studying $\mathcal{PT}$-symmetric QFTs  
is that, for sufficiently small values of $\epsilon$, it allows to perform the functional integration for the calculation of the Green's functions along the real-$\phi$ axis, rather than in the complex-$\phi$ domain.
%, as for sufficiently small values of $\epsilon$ it converges term-by-term in powers of $\epsilon$\,\cite{Bender:2018pbv}.

Clearly, for studying and understanding the potentialities of this expansion, the higher orders need to be considered. 
%while in\cite{Bender:2018pbv} the Green's functions are calculated at $\mathcal{O}(\epsilon)$ only, for a systematic analysis of this expansion in connection with the  $g\phi^2(i \phi)^\epsilon$ theory, higher orders need to be also considered. 
In this respect, we note that even for the corresponding ordinary theory $g\phi^2(\phi^2)^{\epsilon}$, to which the logarithmic expansion was previously applied\cite{Bender:1987dn,Bender:1988rq}, only the $\mathcal O(\epsilon)$ was systematically calculated, while only partial results exist for the $\mathcal{O}(\epsilon^2)$\cite{Bender:1988ux,Bender:1988ig}. 

Now, since the logarithmic expansion and the non-hermitian   $g\phi^2 (i\phi)^{\epsilon}$ theory have each their own peculiarities, in order to better investigate the properties both of the expansion and of the $\mathcal{PT}$-symmetric theory itself, we decided to disentangle the two investigations, splitting our analysis in two different works. The present paper is the second one of this series. 

In the first paper of this series\cite{Branchina:2020jhd}  we studied the higher orders in the expansion of the ordinary theory $g\phi^2 (\phi^2)^{\epsilon}$.
We found that at any finite order in $\epsilon$ (for dimension $d\geq 2$) the theory is non-interacting, i.e. all the Green's function $\mathcal G_n$ with $n \geq 4$ vanish,
%: truncating the interaction term to a finite power of $\log\phi$ (that is to a finite power of  $\epsilon$) gives  a "too poor" approximation to the physical interaction, not sufficient to guarantee the existence of non-trivial $S$-matrix elements. 
and concluded that physically sensible results can be obtained only resorting to resummations. However, resumming leading and subleading contributions to the $\mathcal{G}_n$, we found non-vanishing but quite trivial results. In the particular case of $\epsilon=2$, the ordinary $g\phi^4$ theory, these resummations lead to the first and second order results of the weak-coupling expansion.  

The goal of the present work is to study the Green's functions of the  $\mathcal{PT}$-symmetric  $g\phi^2 (i \phi)^{\epsilon}$ theory  at higher orders in $\epsilon$ (the $\mathcal{O}(\epsilon)$ was considered in\cite{Bender:2018pbv}). For this analysis, we will take great advantage of the techniques that we developed in our previous paper\cite{Branchina:2020jhd}. Interestingly we will see that the results for the $\mathcal{PT}$-symmetric theory closely parallel those that we found for the ordinary theory. 

We begin by calculating the Green's functions at $\mathcal{O}(\epsilon^2)$, giving a closed form for all the $\mathcal{G}_n$ in terms of hypergeometric functions, and find that up to this order the theory is non-interacting. We then move to consider higher orders. The result does not change: {\it at each finite order in $\epsilon$, the theory is non-interacting for any dimension $d \geq 2$}. We then consider resummations of the Green's functions, and find that, as it was the case for the corresponding ordinary theory, the results that we obtain are quite deceptive. Specifying to the $i g \phi^3$ model, we see that the resummations simply give the lowest orders of the weak-coupling expansion. 

The rest of the paper is organized as follows.   
In Section 2 we begin by setting up the tools for our analysis, and review some previous results. 
%of the results of\cite{Bender:2018pbv}. 
We also analyse the UV behaviour of the $\mathcal{O}(\epsilon)$ contributions to the $\mathcal{G}_n$.
%, and show that at this order the theory is non-interacting for dimensions $d\geq2$.
In Section 3 we move to the $\mathcal{O}(\epsilon^2)$, obtaining general expressions for {\it all the Green's functions} $\mathcal G_n$. We study the UV behaviour of the $\mathcal G_n$, and then extend the analysis to higher orders in Section 4.
Section 5 is devoted to the resummation of the UV-leading contributions at each order in $\epsilon$, while in Section 6 we consider the resummation of subleading terms.
Section 7 is for the conclusions.

\section{Green's functions at $O(\epsilon)$ and their UV behaviour}
In the present section, after %setting up the tools for our analysis by 
sketching the steps for the calculation of the $\mathcal{G}_n$ at each order in $\epsilon$, we review the first order calculations, thus obtaining the results of\cite{Bender:2018pbv}, with some generalizations detailed below. We then analyse the UV behaviour of the Green's functions, showing that at this order {\it the theory is non-interacting}.  We also introduce ``effective vertices" that will play a very important role for the analysis of the theory at higher orders.

Let us consider the $\mathcal{PT}$-symmetric (Euclidean) lagrangian in $d$ dimensions ($\epsilon \geq 0$):
\begin{equation} \label{eq:L}
\mathcal{L} = \frac{1}{2}(\nabla\phi)^2+\frac{1}{2}m^2\phi^2+\frac{1}{2}g\mu^2\phi^2\left(i\mu^{1-d/2}\phi\right)^\epsilon\,,
\end{equation}
that is an extension of the lagrangian considered in \cite{Bender:2018pbv}, where we introduce an explicit mass term $m^2\phi^2$, and use dimensionful parameters, with $g$ being the coupling constant and $\mu$ the 't Hooft scale. 
Expanding  $\left(i\mu^{1-d/2}\phi\right)^\epsilon$ in the interaction lagrangian in powers of $\epsilon$,
and including the $\epsilon^0$ term in the ``free lagrangian" (see below), (\ref{eq:L}) can be written as
\begin{equation}\label{splitlagr}
\mathcal{L}=\mathcal{L}_0+ {\mathcal{L}_{int}}\equiv \mathcal{L}_0+\sum_{p=1}^{\infty}\mathcal{L}_p\,, 
\end{equation}
with $\mathcal{L}_0$ and $\mathcal L_p$ given by 
\begin{align} \label{L0} \mathcal{L}_0 &\equiv \frac{1}{2}(\nabla\phi)^2 + \frac{1}{2}\left(m^2+g\mu^2\right)\phi^2 
\\
\label{Lp}
\mathcal{L}_p&\equiv \frac{1}{2} g\mu^2 \, \frac{\epsilon^p}{p!} \, \phi^2\, \log^p\left(i\mu^{1-d/2}\phi\right)\,.
\end{align}

As $\phi$ is real, the complex logarithm is $ \log\left(i\mu^{1-d/2}\phi\right)=\frac{1}{2}\left[\log\left(\mu^{2-d}\phi^2\right) + i\pi\frac{|\phi|}{\phi}\right]\, $, so 
$\mathcal{L}_p$ can be splitted as %(\textcolor{red}{levare da qui ... mettere altrove }$\lambda_p=\frac{1}{2\, p!}g\mu^2\left(\frac{\epsilon}{2}\right)^p$)
\begin{equation}\label{Lpk}
\mathcal{L}_p=  
\frac{g\mu^2}{2\, p!}\left(\frac{\epsilon}{2}\right)^p
\sum_{k=0}^{p}
\binom{p}{k} \,\phi^2 \,\log^{p-k}\left(\mu^{2-d}\phi^2\right) \left(i\pi \frac{|\phi|}{\phi}\right)^{k} \equiv
\sum_{k=0}^{p} \mathcal{L}_{pk}\,\,.
\end{equation}

\begin{comment}
\begin{equation} \label{Lpk}
\mathcal{L}_{pk} = \lambda_p \binom{p}{k} \,\phi^2 \,\log^{p-k}\left(\mu^{2-d}\phi^2\right) \left(i\pi \frac{|\phi|}{\phi}\right)^{k}\,.
\end{equation}
\end{comment}

\begin{comment}
To perform the expansion, we treat $\mathcal L_{int}$ as a perturbation and expand the exponential $e^{-S_{int}}$ in the path integral that defines the connected Green's functions as:
\begin{align} \label{orders}
& e^{-S_{int}} = 1  - \int d^du\,\mathcal{L}_1  +  \Bigg[-\int d^du\,\mathcal{L}_2 +\frac{1}{2}\int d^du\,\mathcal{L}_1 \int d^dw\,\mathcal{L}_1\Bigg] \nonumber \\  & \quad + \Bigg[-\int d^du\,\mathcal{L}_3 +\int d^du\,\mathcal{L}_1 \int d^dw\,\mathcal{L}_2 -\frac{1}{6}\int d^du\,\mathcal{L}_1\int d^dw\,\mathcal{L}_1\int d^dz\,\mathcal{L}_1\Bigg] +  \ldots 
\end{align}
i.e. collecting the different powers of $\epsilon$.
\end{comment}

The expansion of the Green's functions  in the parameter $\epsilon$ is obtained by treating $\mathcal L_{int}$ as a perturbation, and expanding as usual the exponential $e^{-S_{int}}$ in the path integral that defines the ${\mathcal G}_n$. We will consider only connected Green's functions.

From (\ref{Lpk}) we see that all the ``interaction terms" $\mathcal{L}_{pk}$ contain factors of the kinds $\phi^2\log^k (\mu^{2-d}\phi^2)$ and $i\pi |\phi|/\phi$. Therefore, in addition to the terms that appear in the corresponding ordinary theory $g\phi^2(\phi^2)^{\epsilon}$ considered in\cite{Branchina:2020jhd}, in the present case also powers of $|\phi|/\phi$ appear. 
The calculation of the different orders in $\epsilon$ of the connected Green's functions can be done as  sketched below\cite{Bender:2018pbv}. The powers of $\log \phi^2$ are replaced by the formal identity (that is an adaptation of the replica trick\,\cite{Parisi})
\begin{equation}\label{replica}
\mu^{2-d}\phi^2 \log^k (\mu^{2-d}\phi^2) = \lim_{N \rightarrow 1} \frac{d^k}{dN^k} \mu^{(2-d)N}\phi^{2N} \,,
\end{equation}
and the factors $\phi |\phi|$ are written with the help of the integral formula
\begin{equation}\label{mod}
\frac{|\phi|}{\phi}= \frac 2 \pi  \int_0^{\infty}dt\,\frac{\sin (t \phi)}{t}=\frac 2 \pi \int_0^{\infty}dt\,\sum_{r=0}^{\infty}\frac{(-1)^r \, t^{2r}}{(2r+1)!}\, \phi^{2r+1}\,.
\end{equation}

The variables $N_i$ are initially treated as positive integer numbers, so that from both  (\ref{replica}) and (\ref{mod}) the functional integrals for the calculation of the $\mathcal G_n$ are traced back to an application of the Wick's theorem. The results are then analytically extended to real values of the $N_i$'s. %\textcolor{red}{Delicate issues related to this procedure were discussed in detail in \cite{Branchina:2020jhd}, and we will se in the following that similar issues also arise from the application of (\ref{mod}).}

\subsection{Green's functions at $\mathcal{O}(\epsilon)$. Effective vertices.}

To put at work the above machinery, we begin by calculating the $\mathcal O(\epsilon)$ contribution to the connected Green's functions, thus recovering results similar to those of\,\cite{Bender:2018pbv}, but relative to the lagrangian (\ref{eq:L}). 
From\,(\ref{splitlagr}) we have (we need only the free partition function $Z_0$ as we consider  connected $\mathcal{G}_n$):
\begin{equation}
\mathcal{G}_n^{(\epsilon)} (x_1,\dots,x_n) = -\frac{1}{Z_0} \int \mathcal{D}\phi\,\, e^{-S_0}\, \phi(x_1)\dots\phi(x_n) \int d^du\, \mathcal{L}_1[\phi(u)]\,.
\end{equation}

From (\ref{Lpk}) we see that  $\mathcal{L}_1 = \mathcal{L}_{1\,0} + \mathcal{L}_{1\,1}$. Since $\mathcal{L}_{1\,0}$ is even in $\phi$ while $\mathcal{L}_{1\,1}$ is odd, they contribute to even and odd $\mathcal{G}_n$ respectively. We treat them separately.
\vskip10pt

{\bf Odd Green's functions.} %$\mathcal{G}_n$.}
The $\mathcal{\mathcal{O}(\epsilon)}$ contributions to the odd Green's functions come from $\mathcal{L}_{1\,1}$.
%, whose presence is due to the non-hermiticity of the theory.
From (\ref{Lpk}) and (\ref{mod}) we then have:
\begin{align} \label{Gnodd}
&\mathcal{G}_n^{(\epsilon)} (x_1,\dots,x_n) 
%&= -\frac{1}{Z_0} \int D\phi\,\, e^{-S_0}\, \phi(x_1)\dots\phi(x_n) \int d^du\, \mathcal{L}_{1\,1}[\phi(u)] \nonumber \\
= -\frac{1}{Z_0} \int \mathcal{D}\phi\,\, e^{-S_0}\, \phi(x_1)\dots\phi(x_n) \int d^du\, \frac{\epsilon}{4}g\mu^2 \, i \pi |\phi(u)|\phi(u) \nonumber\\
&=- i\, \frac{\epsilon}{2}g\mu^2 \int d^du \int_{0}^{\infty} dt\, \sum_{r=0}^{\infty}\frac{(-1)^r \, t^{2r}}{(2r+1)!}\, \frac{1}{Z_0}\int \mathcal{D}\phi\, e^{-S_0}\, \phi(x_1)\dots\phi(x_n)\, \phi^{2r+3}(u)\,.
\end{align}
For $r<\frac{n-3}{2}$ the functional integral vanishes, as we cannot draw connected diagrams, while for $r\geq \frac{n-3}{2}$ we get
\begin{equation}\label{oddpathint}
\frac{1}{Z_0}\int \mathcal{D}\phi\, e^{-S_0}\, \phi(x_1)\dots\phi(x_n)\, \phi^{2r+3}(u) = B_n(r) \Delta(0)^{r-\frac{n-3}{2}} \prod_{i=1}^{n} \Delta(x_i-u)
\end{equation}
where $B_n(r)$ is the combinatorial factor coming from the different contactions
\begin{equation}\label{B}
B_n(r)=(2r+3)\dots(2r+4-n)(2r+2-n)!!= 2^{\frac{n-1}{2}}\, (r+1)_{\frac{n-1}{2}} \,(2r+3)!! \,,
\end{equation}
and $\Delta(x-u)$ is the free propagator ($M^2\equiv m^2+g \mu^2$)
\begin{eqnarray}\label{propagator}
\Delta(x-u) =  \int \frac{d^dp}{(2\pi)^d}\,\frac{1}{p^2+M^2}\,e^{-ip\cdot(x-u)} \quad \to \quad
\Delta(0) =  \int \frac{d^dp}{(2\pi)^d}\,\frac{1}{p^2+M^2}\,\,.
\end{eqnarray}

The loop integral $\Delta(0)$ diverges for $d \geq 2$, and for the purposes of our later analysis we will regularize this divergence with the help of a physical momentum cut-off, that is
\begin{equation} \label{delta0UV}
	\Delta(0) \sim 
	\begin{cases}
	\Lambda^{d-2} \qquad &{\rm for}\,d>2\\
	\log \Lambda \qquad &{\rm for}\,d=2
	\end{cases}\,.
\end{equation}

Due to the presence of the falling factorial $(r+1)_{\frac{n-1}{2}}$, the coefficient $B_n(r)$ vanishes for $r<\frac{n-3}{2}$,  so that (\ref{oddpathint}) holds true for all positive integers $r$. 
Therefore, the $\mathcal{\mathcal{O}(\epsilon)}$ contribution (\ref{Gnodd}) to the odd Green's functions can be written as
\begin{equation} \label{Gn1odd}
\mathcal{G}_n^{(\epsilon)} (x_1,\dots,x_n) = \Pi^{(\epsilon)}_n \int d^du\, \prod_{i=1}^{n} \Delta(x_i-u)\,, 
\end{equation}
where, following our previous  work on the corresponding ordinary theory\cite{Branchina:2020jhd}, we have defined the %$\mathcal{\mathcal{O}(\epsilon)}$
{\it effective vertices}\, $\Pi^{(\epsilon)}_n$ (with an  odd number $n$ of external legs) as
\begin{align}\label{Pi1odddef}
\Pi^{(\epsilon)}_n & \equiv - i\, \frac{\epsilon}{2}g\mu^2 \int_{0}^{\infty} dt\, \sum_{r=0}^{\infty}\frac{(-1)^r \, t^{2r}}{(2r+1)!}\, B_n(r)\, \Delta(0)^{r-\frac{n-3}{2}}\,.
\end{align}
The series over $r$ and the integral over $t$ can be performed, and we finally get:
\begin{equation}\label{Pi1odd}
\Pi_n^{(\epsilon)}=\frac{\epsilon}{2} g\mu^2      (-i)^{n+2}\,\Gamma\left(\frac{n}{2}-1\right)\left[\frac{\Delta(0)}{2}\right]^{1-\frac{n}{2}}  \,.
\end{equation}

Even though (\ref{Pi1odd}) is obtained for odd values of $n$, we will see below that the calculation of even Green's functions will also result in effective vertices $\Pi_n^{(\epsilon)}$ that are again  given by (\ref{Pi1odd}) (with the exception of the special case $n=2$).  

From (\ref{Gn1odd}) we see that the Green's functions  
at $\mathcal{O(\epsilon)}$ are written in terms of coordinate-independent (and then momentum-independent) effective vertices $\Pi_n^{(\epsilon)}$ of order $\epsilon$, to which $n$ external legs are attached. The effective vertex  $\Pi_n^{(\epsilon)}$ is obtained once (at least part of) the interaction is integrated out, and as such it is a dressed vertex function. Later we will also introduce higher order effective vertices $\Pi_n^{(\epsilon^p)}$.  

\vskip 10pt

\begin{comment}
\noindent\textit{One-point Green's function $\mathcal{G}_1$.}
For $n=1$ the falling factorial gives just a factor $1$, so in (\ref{Pi1oddbis}) we simply need to perform the Gaussian integrals, thus getting
\begin{align}
\Pi_1^{(\epsilon)} &= - 2 \,i\, \lambda_1\, \Delta(0) \int_{0}^{\infty} dt\, \sum_{r=0}^{\infty} \frac{1}{r!}\left(-\frac{\Delta(0)\, t^2}{2}\right)^r (2r+3) = - i\, \epsilon\,g\mu^2 \sqrt{\frac{\pi \Delta(0)}{2}}
%&=- 2 \,i\, \lambda_1\, \Delta(0) \int_{0}^{\infty} dt\, \left[-\Delta(0) \,t^2  + 3\right] e^{-\frac{\Delta(0) t^2}{2}} \nonumber \\
%&= -4\,i\, \lambda_1 \sqrt{\frac{\pi \Delta}{2}} = - i\, \epsilon\,g\,\mu^2 \sqrt{\frac{\pi \Delta}{2}}
\end{align}
Then, using $\int d^du\, \Delta(x-u) = \frac{1}{M^2}$, for $\mathcal{G}_1^{(\epsilon)}$ we have
\begin{equation} \label{G1eps}
\mathcal{G}_1^{(\epsilon)}= \Pi_1^{(\epsilon)} \int d^du\, \Delta(x-u) = - i\, \epsilon\frac{g\mu^2}{m^2+g\mu^2} \sqrt{\frac{\pi \Delta(0)}{2}}\,.
\end{equation}
\textit{Odd Green's functions $\mathcal{G}_n$ with $n >1$.} In this case the falling factorial is non trivial, and the terms with $r<\frac{n-3}{2}$ vanish. After simplifying the factorials, in (\ref{Pi1oddbis}) we finally resort to simple Gaussian integrals, thus getting:
\begin{equation}
\int_{0}^{\infty} dt\, \sum_{r=0}^{\infty} \frac{1}{r!}\left(-\frac{\Delta(0)\, t^2}{2}\right)^r (2r+3) (r+1)_{\frac{n-1}{2}} = \frac{i^{n-3}}{\sqrt{2 \Delta(0)}}\, \Gamma\left(\frac{n}{2}-1\right) 
\end{equation}
so that for the $n$-legs effective vertex with odd $n>1$ we obtain:
\end{comment}

{\bf Even Green's functions.} %$\mathcal{G}_n$.}
The $\mathcal{\mathcal{O}(\epsilon)}$ contributions to the even Green's functions come from $\mathcal{L}_{1\,0}$, and coincide with the terms of the same order of the corresponding  $g \phi^{2}(\phi^2)^\epsilon$ hermitian theory \cite{Branchina:2020jhd}. Using the replica trick for the logarithm in $\mathcal{L}_{1\,0}$ we have
\begin{align}
\mathcal{G}_n^{(\epsilon)} (x_1,\dots,x_n) 
&= -\frac{1}{Z_0} \int \mathcal{D}\phi\,\, e^{-S_0}\, \phi(x_1)\dots\phi(x_n) \int d^du\, \frac{\epsilon}{4}g\mu^2\, \phi^2(u) \,\log\left(\mu^{2-d}\phi^2(u)\right) \nonumber\\
&=-\frac{\epsilon}{4}g\mu^2\int d^du \lim_{N\to1} \frac{d}{dN} \, \mu^{(2-d)(N-1)} \frac{1}{Z_0}\int \mathcal{D}\phi\, e^{-S_0}\, \phi(x_1)\dots\phi(x_n)\,\phi^{2N}(u)\,.
\end{align}

As explained before, the functional integral is calculated by considering first integer values of $N$. For $N<\frac{n}{2}$\, the path integral vanishes, as we cannot draw connected diagrams, while for $N\geq \frac{n}{2}$ we have
\begin{equation} \label{Evenpathint}
\frac{1}{Z_0}\int \mathcal{D}\phi\, e^{-S_0}\, \phi(x_1)\dots\phi(x_n)\,\phi^{2N}(u) = C_n(N) \Delta(0)^{N-\frac{n}{2}} \prod_{i=1}^{n} \Delta(x_i-u)\,,
\end{equation}
where $C_n(N)$ is the combinatorial factor coming from the contractions:
\begin{equation}\label{C}
C_n(N)=2N(2N-1)\dots(2N-n+1) (2N-n-1)!! = 2^{\frac{n}{2}} (N)_{\frac{n}{2}} (2N-1)!!\,.
\end{equation}

As for $N<\frac{n}{2}$ the falling factorial $(N)_{\frac{n}{2}}$ vanishes, Eq.\,(\ref{Evenpathint}) holds true even for $N<\frac{n}{2}$, and then for all positive integers $N$. Starting from this observation, the required analytic extension of (\ref{Evenpathint}) to real values of $N$ is obtained with the help of the  
identity $(2N-1)!!= 2^{N-1} \frac{\Gamma(N+\frac{1}{2})}{\Gamma(\frac{3}{2})}$, and this finally allows to take the derivative with respect to $N$ and the limit $N\to1$.

The $\mathcal{\mathcal{O}(\epsilon)}$ contribution to the even Green's functions $\mathcal{G}_n$ is then 
\begin{equation} \label{Gn1even}
\mathcal{G}_n^{(\epsilon)} (x_1,\dots,x_n) = \Pi^{(\epsilon)}_n \int d^du\, \prod_{i=1}^{n} \Delta(x_i-u)\,,
\end{equation}
where we defined the even $n$-legs $\mathcal{\mathcal{O}(\epsilon)}$ effective vertex $\Pi^{(\epsilon)}_n$ as
\begin{align}\label{Pi1evendef}
\Pi^{(\epsilon)}_n &\equiv -\frac{\epsilon}{4}g\mu^2\, \lim_{N\to1} \frac{d}{dN} \left[ \mu^{(2-d)(N-1)}\, C_n(N)\, \Delta(0)^{N-\frac{n}{2}} \right]\,.
\end{align}

Performing the derivative and the limit we finally obtain:
\begin{align} \label{Pi1even2}
&\Pi_2^{(\epsilon)}=
-\frac{\epsilon}{2}g\mu^2\,K  \\
&\Pi_n^{(\epsilon)}=
\frac{\epsilon}{2}g\mu^2(-i)^{n+2}\,\Gamma\left(\frac{n}{2}-1\right)\left[\frac{2}{\Delta(0)}\right]^{\frac{n}{2}-1}  \qquad {\rm for} \, n \neq 2 \label{Pi1even} 
\end{align}
where 
\begin{equation} \label{K}
K\equiv 1+\log\left[2\mu^{2-d}\Delta(0)\right]+\frac{\Gamma'\left(\frac{3}{2}\right)}{\Gamma\left(\frac{3}{2}\right)} \,.
\end{equation}

Comparing (\ref{Pi1even}) with (\ref{Pi1odd}) we see that, as anticipated above, the two expressions for $\Pi_n^{(\epsilon)}$ with $n$ even and odd coincide. 
%From now on we will refer to (\ref{Pi1even}) and/or (\ref{Gn1even}) for both even and odd effective vertices odd Green's functions.

%\textcolor{red}{.. Migliorare commenti .. Although the results for the  $\mathcal{G}_n^{(\epsilon)}$ are essentially those of \cite{Bender:2018pbv}, the important novelty of this section is in the $\Pi^{(\epsilon)}_n$. The introduction of these effective vertices provides a great advancement for several reasons. We will see that they allow to organize the expansion in $\epsilon$ of the $\mathcal G_n$ in a form that becomes crucial when considering higher orders. Moreover from the explicit expression (\ref{Gn1odd}) and (\ref{Gn1even}) for the $\mathcal{G}_n^{\epsilon}$ we see that the behaviour of the Green's function in the UV region can be read directly from the effective vertices. This is what we are going to consider in the coming subsection.}

The introduction of these effective vertices provides a great advancement for several reasons. As for the corresponding ordinary theory\cite{Branchina:2020jhd}, we will see that, at any order in $\epsilon$, the Green's function can be written in terms of effective vertices and loop integrals. Moreover, we will show that the effective vertices are extremely useful for analysing the UV behaviour of the Green's functions at each order in $\epsilon$. In the next subsection we start with the $\mathcal{O}(\epsilon)$.

\subsection{UV behaviour of the Green's functions at $\mathcal{O(\epsilon)}$}

When studying a quantum field theory, it is of the greatest importance to analyse the UV behaviour of the Green's functions. We now perform such an analysis, and this is another novelty of the present section.
To this end, we observe that the loop integral $\Delta(0)$ in (\ref{propagator}), that appears in the $\Pi_n^{(\epsilon)}$, diverges for $d\geq2$. From now on we are interested in studying these cases. 

Starting with the one-point Green's function $\mathcal G_1$, that is nothing but the vacuum expectation value $\mathcal G_1 = \langle 0|\phi|0\rangle = v$, from (\ref{Gn1odd}) for $n=1$ we see that:
\begin{align}\label{UVG1}
\mathcal G_1^{(\epsilon)}  \sim - i\,\epsilon\, \Delta(0)^{\frac 12}\,.
\end{align}

Needless to say, this divergence is trivially renormalized inserting in the lagrangian a linear counterterm  $\delta v \phi$. In this respect, we observe that, differently to what is claimed in\cite{Bender:2018pbv}, having got a \textit{non-vanishing negative imaginary} result (actually negative imaginary \textit{infinite}) for the unrenormalized $\mathcal G_1^{(\epsilon)}$ is totally irrelevant in connection with the expectation that $\mathcal G_1$ (as a result of the $\mathcal{PT}$-symmetric nature of the theory) should be a negative imaginary number. The renormalization can produce any possible imaginary value, including $\mathcal G_1^{ren}=0$. 

Let us move now to the two-point Green's function $\mathcal{G}_2$.
Going to momentum space, and adding to (\ref{Gn1even}) the order $\epsilon^0$ free propagator, up to $\mathcal{O}(\epsilon)$ we have (remember that $M^2=m^2+g \mu^2$)

\begin{equation}\label{G2fourier}
\widetilde{\mathcal{G}}_2(p) = \frac{1}{p^2+M^2} - \frac{1}{p^2+M^2} \left (\frac{\epsilon g\mu^2 K}{2} \right )\frac{1}{p^2+M^2} \,.
\end{equation}
From (\ref{K}) and (\ref{G2fourier}) we see that  the correction to the tree-level result for $\mathcal G_2$ goes as: 
\begin{align}\label{UVG2}
\mathcal G_2 \sim
\log \Delta(0)\,.
\end{align}

The term in the round brackets of (\ref{G2fourier}) is easily recognized as the lowest order 1PI self-energy diagram within the expansion in $\epsilon$. The resummation of the geometric series (in increasing powers of $\epsilon$) can be then performed, thus obtaining for $\widetilde{\mathcal G}_2$:
\begin{equation}\label{G2p}
\widetilde{\mathcal G}_2(p) = \frac{1}{p^2+M^2 + \frac 12 \epsilon g\mu^2\,K}\,.
\end{equation}

Inserting (\ref{K}) in (\ref{G2p}), we see that the radiatively corrected mass $m_R^2$ at this order is:
\begin{equation} \label{mReps}
m_{R}^2=M^2 + \frac{\epsilon g\mu^2}{2}  \left\{\log\left[2\mu^{2-d}\Delta(0)\right]+1+\frac{\Gamma'(\frac{3}{2})}{\Gamma(\frac{3}{2})}\right\}\,,
\end{equation}
that diverges as $\log\Lambda$ for any $d>2$, and as $\log(\log\Lambda)$ for $d=2$. This divergence is trivially cancelled once the mass counterterm contained in $M^2$ is tuned accordingly.

Moreover it is worth to note that as compared to the algebraic divergence encountered at first order of the weak-coupling expansion, we here observe a much milder logarithmic divergence.
Actually we have shown\cite{Branchina:2020jhd} that, when $\mathcal{G}_2$ is calculated at the same order of approximation, this result for $\delta m^2$ is also present in the ordinary theory. This shows that such a mild UV behaviour is peculiar of the expansion in $\epsilon$.

Going now to the $\mathcal{G}_n$ with $n>2$ (actually for $n\neq 2$ the equation below incorporates also the case in (\ref{UVG1}) for  $n=1$), from (\ref{Pi1even}) we see that
\begin{align}\label{Gnvanish}
\mathcal G_n \sim
\Delta(0)^{1-\frac n2}
\end{align}
so that, for $d\geq 2$, at $\mathcal{O}(\epsilon)$ all the $\mathcal G_n$ for $n \geq 3$ vanish. 

At this order then {\it the theory turns out to be non-interacting}, and this sounds as a surprising and disturbing result. This point clearly needs to be further investigated, and to this end we have to consider higher orders in the expansion. In the next section we will analyse systematically the $\mathcal{O}(\epsilon^2)$.

\section{Green's functions at $\mathcal{O}(\epsilon^2)$ and their UV behaviour}
The $\mathcal{O}(\epsilon^2)$ contributions to the Green's functions are given by
\begin{align}
\mathcal{G}_n^{(\epsilon^2)} (x_1,\dots,x_n) &= \frac{1}{Z_0} \int \mathcal{D}\phi\,\, e^{-S_0}\, \phi(x_1)\dots\phi(x_n)\, \times \nonumber\\
&\times \left\{-\int d^du\, \mathcal{L}_2[\phi(u)] + \frac{1}{2} \iint d^du_1d^du_2\,\, \mathcal{L}_1[\phi(u_1)]\, \mathcal{L}_1[\phi(u_2)]\right\}\,.
\end{align}
The $\mathcal{L}_2$ term gives diagrams with one vertex, the other one diagrams with two vertices. We treat these two contributions separately, denoting them  as $\mathcal{G}_n^{(\epsilon^2,1)}$ and $\mathcal{G}_n^{(\epsilon^2,2)}$ respectively.

\subsection{Diagrams with one Effective Vertex}

From (\ref{Lpk}) we know that $\mathcal{L}_2 = \mathcal{L}_{2\,0} + \mathcal{L}_{2\,1} + \mathcal{L}_{2\,2}$. Since $\mathcal{L}_{2\,0}$ and $\mathcal{L}_{2\,2}$ are even in $\phi$, while $\mathcal{L}_{2\,1}$ is odd, they contribute only to the even and to the odd Green's functions respectively, so that it is useful to treat the two cases separately.

\paragraph{Odd Green's functions.}
Expliciting  $\mathcal{L}_{2\,1}$ from (\ref{Lpk}), and using (\ref{replica}) and (\ref{mod}), we have
\begin{align}\label{Gnovd}
&\mathcal{G}_n^{(\epsilon^2,1)}(x_1, \dots, x_n)=-\frac{1}{Z_0} \int \mathcal{D}\phi\, e^{-S_0} \phi(x_1)\dots\phi(x_n) \int d^du\, \frac{\epsilon^2}{8}g\mu^2 \,\phi^2(u) \log\left(\mu^{2-d}\phi^2(u)\right)\, i\pi\frac{|\phi(u)|}{\phi(u)}\nonumber\\
&=- i \frac{\epsilon^2}{4}g\mu^2 \int d^d u \,  \lim_{N\to 1}  \frac{d}{dN} \mu^{(2-d)(N-1)} \int_0^\infty dt \sum_{r=0}^{\infty} \frac{(-t^2)^r}{(2r+1)!} \frac{1}{Z_0} \int \mathcal{D}\phi\, e^{-S_0} \phi(x_1)\dots \phi(x_n)\,\phi(u)^{2N+2r+1}\nonumber\\
&=- i \frac{\epsilon^2}{4}g\mu^2 \left[\frac{2}{\Delta(0)}\right]^{\frac{n-1}{2}} \lim_{N\to 1} \frac{d}{dN} \mu^{(2-d)(N-1)} \int_0^\infty dt \sum_{r=0}^{\infty} \frac{(-t^2)^r}{(2r+1)!} \Delta(0)^{r+N} (r+N)_{\frac{n-1}{2}}(2N+2r+1)!! \nonumber\\
&\times \int d^du \prod_{i=1}^{n} \Delta(x_i-u) \,. 
\end{align}

\paragraph{Even Green's functions.} 
The term $\mathcal{L}_{2\,2}$ is quadratic in $\phi$, and then contributes only to $\mathcal{G}_2$, the calculation being trivial. Concerning  the contribution of $\mathcal{L}_{2\,0}$, from (\ref{Lpk}) and (\ref{replica}) we get:
\begin{align}\label{Gnovp}
&\mathcal{G}^{(\epsilon^2,1)}_n(x_1,\dots,x_n) = -\frac{1}{Z_0} \int \mathcal{D}\phi\, e^{-S_0} \phi(x_1)\dots\phi(x_n) \int d^du\,\frac{\epsilon^2}{16}g\mu^2 \,\phi^2(u) \log^2\left(\mu^{2-d}\phi^2(u)\right)\, \nonumber\\
&=-\frac{\epsilon^2}{16}g\mu^2\, \int d^du
\lim_{N \to 1} \frac{d^2}{dN^2}\, \left\{  \mu^{(2-d)(N-1)} \frac{1}{Z_0}\int \mathcal D\phi \,e^{-S_0} \phi(x_1)\ldots\phi(x_n) \phi(u)^{2N} \right\} \nonumber
\\
&= -\frac{\epsilon^2}{16}g\mu^2\,\lim_{N \rightarrow 1} \frac{d^2}{dN^2} \left\{ \mu^{(2-d)(N-1)} \left[\Delta(0)\right]^{N-\frac{n}{2}} C_n(N)\right\}\int d^du \prod_{i=1}^{n}\Delta(x_i-u)\,.
\end{align} 

Performing in (\ref{Gnovd}) and (\ref{Gnovp}) the derivative with respect to $N$, the limit $N \to 1$, the sum over $r$ and the integral over $t$, for generic $n$ (either even or odd)  we get
\begin{equation} \label{Gneps21}
\mathcal{G}_n^{(\epsilon^2,1)}(x_1, \dots, x_n)= \Pi^{(\epsilon^2)}_n \int d^du \prod_{i=1}^{n}\Delta(x_i-u)\,,
\end{equation} 
where we have defined the $\mathcal{O}(\epsilon^2)$ effective vertices
\begin{align}\label{Pi22}
\Pi_2^{(\epsilon^2)}&=
-\frac{\epsilon^2}{8} g\mu^2 \left[K^2 - 1 + \psi'\left(\frac{3}{2}\right)+\pi^2\right] 
\\
\label{Pi2n}
\Pi_n^{(\epsilon^2)}&=\frac{\epsilon^2}{4}g \mu^2 (-i)^{n+2}\, \Gamma\left(\frac n2 -1\right)\,  \left[\frac{2}{\Delta(0)}\right]^{\frac{n}{2}-1}  \left(K- H_{\frac n2 -2}\right) \qquad {\rm for} \, n \neq 2\,,
\end{align}
and $K$ is given in (\ref{K}), while $H_n$ stands for the $n$-th Harmonic number. In the expression for $n=2$, we have also included the contribution from $\mathcal{L}_{2\,2}$.

\begin{comment}
\textcolor{red}{Come nel caso dell'ordine $\epsilon$ che si scrive come un $\Pi$ per i propagatori.}

A questo punto sappiamo che fino all'ordine $\epsilon^2$ possiamo scrivere:
\begin{equation} \label{Gneps2fin}
\mathcal{G}_n(x_1, \dots, x_n)=[ \Pi^{(\epsilon)}_n + \Pi^{(\epsilon^2)}_n] \int d^du \prod_{i=1}^{n}\Delta(x_i-u)\,,
\end{equation}
Bellezza dei $\Pi$
\end{comment}

\paragraph{Analysis of the UV behaviour.}
Eq.\,(\ref{Gneps21}) shows that the UV behaviour of the generic $\mathcal{G}_n^{(\epsilon^2,1)}$ is entirely given by the UV behaviour of the $\Pi_n^{(\epsilon^2)}$ in (\ref{Pi22}) and (\ref{Pi2n}), so that:
\begin{align}\label{G2vanish2}
\mathcal{G}^{(\epsilon^2,1)}_{2} &\sim 
\log^2 \Delta(0) \\
\label{Gnvanish2}
\mathcal{G}^{(\epsilon^2,1)}_{n} &\sim \Delta(0)^{1-\frac n2} \log \Delta(0) \qquad {\rm for}\, n \neq 2\,.
\end{align}

Comparing (\ref{G2vanish2}) and (\ref{Gnvanish2}) with the behaviour of the $\mathcal{G}_n$ at $\mathcal{O}(\epsilon)$ (Eqs.\,(\ref{UVG2}) and (\ref{Gnvanish})), we see that the leading behaviour of this one-vertex contribution is enhanced only by a power of $\log\Lambda$ ($d>2$) or $\log\left(\log\Lambda\right)$ ($d=2$). We note that  $\mathcal{G}_1$ and $\mathcal{G}_2$ get a divergent contribution, while   
$\mathcal{G}_n^{(\epsilon^2,1)}$ vanishes when $n\geq 3$. 

However, 
before drawing any conclusion, we have to consider the other $\mathcal{O}(\epsilon^2)$  diagrams, namely the diagrams with two vertices. We do that in the next subsection.  

\subsection{Diagrams with two Effective Vertices}

At $\mathcal{O}(\epsilon^2)$,
the diagrams with two vertices  come from $\mathcal{L}_1 \cdot \mathcal{L}_1$ terms.  
From (\ref{Lpk}) we see that $\mathcal{L}_1= \mathcal{L}_{1\,0} + \mathcal{L}_{1\,2}$, and then the product $\mathcal{L}_1 \cdot \mathcal{L}_1$ splits in the sum of three contributions,  $\mathcal{L}_{1\,0}\cdot\mathcal{L}_{1\,0} + 2\,\mathcal{L}_{1\,0}\cdot\mathcal{L}_{1\,1}+\mathcal{L}_{1\,1}\cdot\mathcal{L}_{1\,1}$  (the mixed term gets a factor two as there are two contributions that are equivalent under the exchange $u_1\iff u_2$). The $\mathcal{L}_{1\,0}\cdot\mathcal{L}_{1\,1}$ term is odd in $\phi$, and gives non-vanishing contributions only to the odd Green's functions, while the first and the third terms are even, and contribute only to the even Green's functions. Let us treat the two cases separately.

\paragraph{Odd Green's functions.} Taking from (\ref{Lpk}) the expression for $ \mathcal{L}_{1\,0}\cdot\mathcal{L}_{1\,1}$, and applying (\ref{replica}) and (\ref{mod}), we obtain:
\begin{align}\label{Gn2odd}
\mathcal{G}_{n}^{(\epsilon^2,2)}(x_1, \dots, x_n)
&=\frac{1}{Z_0} \int \mathcal{D}\phi\,\, e^{-S_0}\, \phi(x_1)\dots\phi(x_n) \times \nonumber\\
&\times \left(\frac{\epsilon}{4}g\mu^2\right)^2 \,i\pi\int d^du_1d^du_2\, \phi^2(u_1) \log\left(\mu^{2-d}\phi^2(u_1)\right) \phi(u_2)|\phi(u_2)|\nonumber\\
&= i \frac{\epsilon^2}{8}g^2\mu^4 \int d^du_1 \,d^du_2\,  \lim_{N\to 1} \frac{d}{dN} \mu^{(2-d)(N-1)} \int_0^\infty dt \sum_{r=0}^{\infty} \frac{(-t^2)^r}{(2r+1)!}
\nonumber\\
&\times\frac{1}{Z_0} \int \mathcal{D}\phi\, \phi(x_1)\dots \phi(x_n) e^{-S_0}\phi(u_1)^{2N}\phi(u_2)^{2r+3}\,. 
\end{align}

Let us evaluate the path integral in (\ref{Gn2odd}), starting by contracting the first $s\, (\leq n)$ fields $\phi(x_i)$ with $\phi(u_1)$ and the remaining $n-s$ fields with $\phi(u_2)$. The other diagrams are obtained by permutations of the external legs. We have to distinguish the case with an even number of internal lines, corresponding to an even number of external legs attached to the vertex $u_1$ ($s=2j$), from the case with an odd number of external legs attached to $u_1$  ($s=2j+1$). These contributions will be denoted with $\mathcal{G}_{n,\,A}^{(\epsilon^2,2)}$ and   $\mathcal{G}_{n,\,B}^{(\epsilon^2,2)}$  respectively.  
%($0 \leq j \leq \frac{n-1}{2}$) .

Let us start with the $A$-type contribution. The connected diagrams that can be drawn from the path integral are of the kind ($2l$ is the even number of internal lines)
	
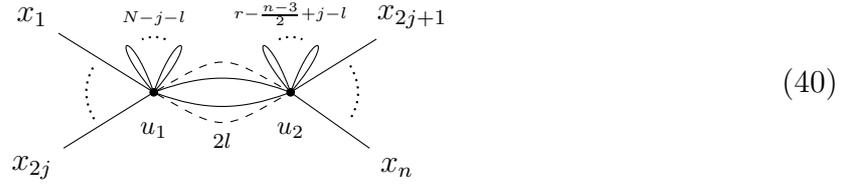
\begin{equation} \label{diagG22nOA}
	\begin{tikzpicture}[baseline=(m)]
	\begin{feynman}[inline=(m)]
	\vertex[dot] (m) at (-0.9, 0) {};
	\vertex[label={center:{\footnotesize $u_1$}}] (m0) at (-0.9,-0.5) {};
	\vertex[dot] (n) at (0.9, 0) {};
	\vertex[label={center:{\footnotesize $u_2$}}] (n0) at (0.9,-0.5) {};
	\vertex (a) at (-2.5,1) {$x_{1}$};
	\vertex (b) at (-2.5,-1) {$x_{2j}$};
	\vertex (c) at (2.5,1) {$x_{2j+1}$};
	\vertex (d) at (2.3,-1) {$x_{n}$};
	\vertex (l1) at (1.6,0.50) {};
	\vertex (l2) at (1.6,-0.50) {};
	\vertex (p1) at (-1.6,0.5) {};
	\vertex (p2) at (-1.6,-0.5) {};
	\vertex (q1) at (-1.2,0.65) {};
	\vertex (q2) at (-0.6,0.65) {};
	\vertex (t1) at (1.2,0.65) {};
	\vertex (t2) at (0.6,0.65) {};
	\diagram* {
		(a) -- (m) -- [out=70, in=50, loop, min distance=0.9cm] m -- [out=130, in=110, loop, min distance=0.9cm] m -- (b);
		(m) -- [half left,out=30, in=150, dashed](n);
		(m) -- [half left,out=-30, in=-150, edge label'={\scriptsize $2l$}, dashed](n);
		(m) -- [out=20,in=160](n);
		(m) -- [out=-20,in=-160](n);
		(c) -- (n) -- [out=70, in=50, loop, min distance=0.9cm] n -- [out=130, in=110, loop, min distance=0.9cm] n -- (d);	
		(l1) -- [thick,dotted,out=-60, in=60](l2);
		(p2) -- [thick,dotted,out=120, in=-120](p1);
		(q2) -- [thick,dotted,out=160, in=20, edge label'={\tiny $N\!-j\!-l$}](q1);
		(t2) -- [thick,dotted,out=20, in=160, edge label= {\tiny $r\!-\frac{n-3}{2}\!+j\!-l$}](t1);
	};
	\end{feynman}
	\end{tikzpicture}
	\end{equation}
Note that $0\leq j\leq \frac{n-1}{2}$. Moreover 
for generic integers $r\geq0$ and $N>0$, connected diagrams are obtained only for  $l\geq1$, and for:
	\begin{equation} \label{conditions}
	\begin{cases}
	N\geq j+l \\
	r\geq \frac{n-3}{2}-j+l
	\end{cases}
	\end{equation}
from which we also have that the number of internal lines is limited by
\begin{equation}
l\leq Min[N-j,\,r-\frac{n-3}{2}+j] \equiv min\,.
\end{equation}	
Under these conditions, the path integral in (\ref{Gn2odd}), i.e. the sum of diagrams of the kind (\ref{diagG22nOA}), gives

\begin{equation}\label{path}
\sum_{j=0}^{\frac{n-1}{2}}\sum_{l=1}^{min}C_{njl}(N,r) \,\Delta(0)^{N-j-l}\, \Delta(0)^{r-\frac{n-3}{2}+j-l}\, \Delta(u_1-u_2)^{2l} \prod_{i=1}^{2j} \Delta(x_i-u_1) \prod_{k=2j+1}^{n} \Delta(x_k-u_2)\,+\,\text{perm.}
\end{equation}
where $C_{njl}(N,r)$ is the combinatorial factor coming from the contractions:
\begin{align}\label{combin}
C_{njl}(N,r) &= (2N)\dots(2N-2j-2l+1)(2N-2j-2l-1)!! \times \nonumber\\
&\times(2r+3)\dots(2r+4-n+2j-2l)(2r+2-n+2j-2l)!! \frac{1}{(2l)!}\nonumber \\
&= \left[(2N-1)!! (N)_{j+l}\, 2^{j+l}\right]\left[(2r+3)!! (r+1)_{\frac{n-1}{2}-j+l}\, 2^{\frac{n-1}{2}-j+l}\right]\frac{1}{(2l)!} \nonumber\\
&= C_{2j+2l}(N)\,B_{n-2j+2l}(r) \,\frac{1}{(2l)!}\,.
\end{align}

In (\ref{combin}) the terms  
$C_{2j+2l}(N)$ and $B_{n-2j+2l}(r)$ are the combinatorial coefficients found at $\mathcal{O}(\epsilon)$, when we were dealing with one-vertex diagrams with even and odd number of legs respectively. Concerning the conditions (\ref{conditions}), it is important to make the following observations.  Due to the presence of the falling factorial, the factor $C_{2j+2l}(N)$, defined in (\ref{C}), vanishes when the condition (\ref{conditions})$_1$ is not satisfied. We can then consider also values of $N<l+j$, and this allows to reach the value $N=1$, that is necessary for the subsequent analytic extension. 
Moreover
the factor $B_{n-2j+2l}(r)$, defined in (\ref{B}), vanishes when $0\leq r\leq \frac{n-1}{2}-j+l $, so that, despite the condition (\ref{conditions})$_2$,  we can take the sum over $r$ in (\ref{Gn2odd}) starting from $r=0$. 
Finally, thanks to these two latter  observations, the sum over $l$ in (\ref{path}) can be extended up to infinity, as (\ref{combin}) vanishes when $l> min$. Then we can write 
\begin{align}
\mathcal{G}^{(\epsilon^2,2)}_{n,\,A} &= \sum_{j=0}^{\frac{n-1}{2}} \int d^du_1 d^du_2 \sum_{l=1}^{\infty} \left[- \frac{\epsilon}{4} g\mu^2 \lim_{N \to 1} \frac{d}{dN} \mu^{(2-d)(N-1)}C_{2j+2l}(N) \Delta(0)^{N-j-l} \right] \nonumber \\ 
&\left[-i\frac{\epsilon}{2} g\mu^2 \int_{0}^{\infty} dt \sum_{r=0}^{\infty} \frac{(-1)^{r} t^{2r}}{(2r+1)!} B_{n-2j+2l}(r)\Delta(0)^{r-\frac{n-3}{2}+j-l}\right] \frac{\Delta(u_1-u_2)^{2l}}{(2l)!} \nonumber \\
&\prod_{i=1}^{2j} \Delta(x_i-u_1) \prod_{k=2j+1}^{n} \Delta(x_k-u_2) + \binom{n}{2j}-1\, \text{perm.}
\end{align}
We easily recognize in the two square brackets the $\mathcal{O}(\epsilon)$ effective vertices $\Pi_k^{(\epsilon)}$ with even and odd number of legs, Eqs.\,(\ref{Pi1evendef}) and (\ref{Pi1odddef}) respectively,  so that
\begin{align} \label{G22oddnAPi}
\mathcal{G}^{(\epsilon^2,2)}_{n,\,A} &= \sum_{j=0}^{\frac{n-1}{2}} \int d^du_1 d^du_2 \sum_{l=1}^{\infty} \Pi^{(\epsilon)}_{2j+2l} \Pi^{(\epsilon)}_{n-2j+2l} \frac{\Delta(u_1-u_2)^{2l}}{(2l)!}\times \nonumber \\
&\times \prod_{i=1}^{2j} \Delta(x_i-u_1) \prod_{k=2j+1}^{n} \Delta(x_k-u_2) + \binom{n}{2j}-1\,\, \text{perm.}
\end{align}

Eq.\,(\ref{G22oddnAPi}) is a compact and elegant form for the $A$-type contribution, written in terms of diagrams built with couples of effective vertices.
Actually we will see that, at each order in $\epsilon$, the $\mathcal G_n$ are all written in terms of effective vertices (those of higher order in $\epsilon$ will be defined later) and loop integrals connecting them. 
This shows the great advantage induced by the effective vertices.

The series over $l$ can be  easily resummed, however we have to distinguish the case $n=1$ from the others. Starting with the former we get:
\begin{align} \label{G221A}
\mathcal{G}^{(\epsilon^2,2)}_{1,\,A} &= \frac{i\,\sqrt{\pi}\,\epsilon^2 g^2 \mu^4}{12}  \left[ \frac{2}{\Delta(0)}\right]^{-\frac 32}  \int d^du_1 \, d^du_2\, \left[ 6\, K \,z - z^2 \,{}_3F_2 \left(1,1,\frac 32;\frac 52,3;z\right)\right] \Delta(x_1-u_1)
\end{align}
while for the others odd values of $n$:
\begin{align} 
\mathcal{G}^{(\epsilon^2,2)}_{n,\,A} &= \frac{\epsilon^2}{2}g^2 \mu^4 (-i)^{n+2} \left[ \frac{2}{\Delta(0)}\right]^{\frac n2-2}\Gamma\left(\frac{n}{2}\right) \nonumber\\
&\times \int d^du_1 \, d^du_2\, \left[K\, z \, - \frac{n}{6}\, z^2 \,{}_3F_2 \left(1,1,1+\frac n2;\frac 52,3;z\right) \right] \prod_{i=1}^{n}\Delta(x_i-u_1)+\nonumber \end{align}
\begin{align}\label{G22noddA}
&+(-i)^n \frac {\epsilon^2}{2}g^2 \mu^4 \left[ \frac{2}{\Delta(0)}\right]^{\frac n2-2}\Gamma\left(\frac{n}{2}-1\right) \nonumber\\
&\times \int d^du_1 \, d^du_2\,\, z\, \,{}_3F_2 \left(1,1,\frac n2-1;\frac 32,2;z\right) \prod_{i=1}^{2}\Delta(x_i-u_1) \prod_{h=3}^n \!\Delta(x_h-u_2)\,\, + {\rm perm.}\,\nonumber\\
&+(-i)^n \frac {\epsilon^2}{4}g^2 \mu^4   \left[ \frac{2}{\Delta(0)}\right]^{\frac n2-2} \sum_{j=2}^{\frac{n-1}{2}} \Gamma (j-1) \Gamma \left(\frac{n}{2}-j-1\right)  \nonumber\\
&\times \int d^du_1 \, d^du_2\, \left[\, _2F_1\left(j-1,\frac{n}{2}-j-1;\frac{1}{2};z\right)-1\right] \prod_{i=1}^{2j}\Delta(x_i-u_1) \prod_{h=2j+1}^n \!\Delta(x_h-u_2)\,\, + {\rm perm.}
\end{align}
where the hypergeometric functions ${}_2 F_{1}$ and ${}_3 F_{2}$ are defined by
\begin{align}
{}_2 F_{1} (a_1,a_2; b; z) &= \sum_{l=0}^{\infty} \frac{a_1^{(l)}a_2^{(l)}}{b^{(l)}} \frac{z^l}{l!}\\
{}_3 F_{2} (a_1,a_2,a_3; b_1,b_2; z) &= \sum_{l=0}^{\infty} \frac{a_1^{(l)}a_2^{(l)}a_3^{(l)}}{b_1^{(l)}b_2^{(l)}} \frac{z^l}{l!}
\end{align}
and we denoted with $z$:
\begin{equation}
z\equiv\left[\frac{\Delta(u_1-u_2)}{\Delta(0)}\right]^2\,.
\end{equation}

Eqs.\,(\ref{G221A}) and (\ref{G22noddA}) show that the series over $l$  give rise to hypergeometric functions that depend on the spacetime integration variables $u_1$ and $u_2$. In this respect, we note that, as long as $u_1\neq u_2$, we have $|z|<1$, and the convergence of the hypergeometric functions is guaranteed. However, when $u_1=u_2$, i.e. $z=1$, convergence issues could arise. More precisely, for ${}_2 F_{1}$ the convergence is guaranteed if
\begin{equation}\label{condhyper1}
	Re(b-a_1-a_2)>0
\end{equation}
while for ${}_3 F_{2}$ the condition is
\begin{equation}\label{condhyper2}
Re(b_1+b_2-a_1-a_2-a_3)>0\,.
\end{equation}

Considering the arguments of the hypergeometric functions in (\ref{G221A}) and (\ref{G22noddA}) we see that both conditions  (\ref{condhyper1}) and (\ref{condhyper2}) reduce to
\begin{equation}
n<5\,.
\end{equation}
This means that, while $\mathcal{G}^{(\epsilon^2,2)}_{1,\,A}$ and $\mathcal{G}^{(\epsilon^2,2)}_{3,\,A}$ are certainly well-behaved, for $n\geq 5$  we need to carefully analyse the singularities in $\mathcal{G}^{(\epsilon^2,2)}_{n,\,A}$ induced by the limit $(u_1-u_2)\to0$, and to investigate whether they are integrable or not. 

Performing the explicit calculation for the cases $d=3$ and $d=4$, we have verified that starting from $\mathcal{G}^{(\epsilon^2,2)}_{5,\,A}$ more and more severe non-integrable singularities in the limit $(u_1-u_2) \to 0$ do actually appear. 
However, when the $B$-type contribution, i.e. diagrams with an odd number of internal lines, is also considered (see below), it turns out that the divergences introduced by these ``$A$-diagrams" are exactly cancelled by opposite divergences due to the ``$B$-diagrams". Similar cancellations also occur for the even Green's functions that we will consider later. For this reason, we do not need  to consider any longer the singularities arising from the hypergeometric functions. 

Let us move now to the $B$-type contribution. The connected diagrams that can be drawn are of the kind
\begin{equation}
\begin{tikzpicture}[baseline=(m)]
\begin{feynman}[inline=(m)]
\vertex[dot] (m) at (-0.9, 0) {};
\vertex[label={center:{\footnotesize $u_1$}}] (m0) at (-0.9,-0.5) {};
\vertex[dot] (n) at (0.9, 0) {};
\vertex[label={center:{\footnotesize $u_2$}}] (n0) at (0.9,-0.5) {};
\vertex (a) at (-2.5,1) {$x_{1}$};
\vertex (b) at (-2.5,-1) {$x_{2j+1}$};
\vertex (c) at (2.5,1) {$x_{2j+2}$};
\vertex (d) at (2.3,-1) {$x_{n}$};
\vertex (l1) at (1.6,0.50) {};
\vertex (l2) at (1.6,-0.50) {};
\vertex (p1) at (-1.6,0.5) {};
\vertex (p2) at (-1.6,-0.5) {};
\vertex (q1) at (-1.2,0.65) {};
\vertex (q2) at (-0.6,0.65) {};
\vertex (t1) at (1.2,0.65) {};
\vertex (t2) at (0.6,0.65) {};
\diagram* {
	(a) -- (m) -- [out=70, in=50, loop, min distance=0.9cm] m -- [out=130, in=110, loop, min distance=0.9cm] m -- (b);
	(m) -- [half left,out=30, in=150, dashed](n);
	(m) -- [half left,out=-30, in=-150, edge label'={\scriptsize $2l+1$}, dashed](n);
	(m) -- (n);
	(c) -- (n) -- [out=70, in=50, loop, min distance=0.9cm] n -- [out=130, in=110, loop, min distance=0.9cm] n -- (d);	
	(l1) -- [thick,dotted,out=-60, in=60](l2);
	(p2) -- [thick,dotted,out=120, in=-120](p1);
	(q2) -- [thick,dotted,out=160, in=20, edge label'={\tiny $N\!-j\!-l\!-1$}](q1);
	(t2) -- [thick,dotted,out=20, in=160, edge label= {\tiny $r\!-\frac{n-3}{2}\!+j\!-l$}](t1);
};
\end{feynman}
\end{tikzpicture}
\end{equation}
Following the same steps made for the $A$-type contribution, we obtain as before a final expression in terms of diagrams built with two $\mathcal{\mathcal{O}(\epsilon)}$ effective vertices $\Pi_k^{(\epsilon)}$, one with even and one with odd number of legs:
\begin{align} \label{G22oddnBPi}
\mathcal{G}^{(\epsilon^2,2)}_{n,\,B} &=  \sum_{j=0}^{\frac{n-1}{2}} \int d^du_1 d^du_2 \sum_{l=0}^{\infty} \Pi^{(\epsilon)}_{2j+2l+2} \Pi^{(\epsilon)}_{n-2j+2l} \frac{\Delta(u_1-u_2)^{2l+1}}{(2l+1)!}\times \nonumber \\
&\times \prod_{i=1}^{2j+1} \Delta(x_i-u_1) \prod_{k=2j+2}^{n} \Delta(x_k-u_2) + \binom{n}{2j+1}-1\,\, \text{perm.}
\end{align}

\begin{comment}
We can perform $2l+1$ ($l \geq 0$) contractions between $\phi(u)$ and $\phi(w)$. Then we obtain
\begin{align}\label{pathintegral2}
&\frac{1}{Z_0} \int \mathcal{D}\phi \phi(x_1)\dots \phi(x_n) e^{-S_0}\phi(u)^{2N}\phi(w)^{2r+3}\nonumber\\ 
&=\frac 12 \sum_{j=0}^{\frac{n-1}{2}} \prod_{i=1}^{2j+1}\Delta(x_i-u) \prod_{h=2j+2}^n \!\Delta(x_h-w)\,2^{\frac{n-1}{2}} \Delta(0)^{N+r-\frac{n-3}{2}} \nonumber\\
&\times (2N-1)!!\,(2r+3)!!\sum_{l=0}^{Min}z^{l+\frac 12 }\frac{2^{2l+1}}{(2l+1)!} (N)_{j+l+1}\,(r+1)_{l-j+\frac{n-1}{2}}\,,
\end{align}
where $Min\equiv \min \left(N-j-1,r+j-\frac{n-3}{2}\right)$ and $z$ defined in (\ref{z}). As in the previous case, we substitute $Min \to \infty$ and we have to study the convergence of the series over $l$. Using the identity $(2l+1)!=2^{2l}l! \left(\frac 32\right)^{(l)}$, we reduce the $l$-dependent part to an hypergeometric function ${}_2F_1$ that always converges for $|z|<1$, while for $|z|=1$ converges only for $n<5$. Then again we ensure the convergence for $n\geq 5$ using a numerical cut-off $L_{max}$, and following the same steps we can write the $\mathcal{G}_n$ in terms of the $\Pi^{(1)}$ in (\ref{Pi1n}):

\begin{align}\label{Gnodd2v2}
&\mathcal{G}^{(\epsilon^2)}_{n}= \sum_{j=0}^{\frac{n-1}{2}} \sum_{l=0}^{L_{max}} \,\,\,\Pi^{(1)}_{2j+2l+2} \,\,\, \Pi^{(1)}_{n-2j+2l}\,\,\frac{1}{(2l+1)!}\nonumber\\
&\times \left[\int d^du \, d^dw\, \prod_{i=1}^{2j+1}\Delta(x_i-u) \prod_{h=2j+2}^n \!\Delta(x_h-w) \,\, \Delta(u-w)^{2l+1}\,\, + \, \binom{n}{2j+1} -1\, {\rm perm.}\right]\,.
\end{align}
\end{comment}

The series over $l$ can be easily resummed. However, as before, we have to distinguish the case $n=1$ from the others. Starting with the former we get:
\begin{align}\label{G221B}
\mathcal{G}^{(\epsilon^2,2)}_{1,\,B} &=\frac{i\, \sqrt{\pi}\, \epsilon^2 g^2 \mu^4}{6\sqrt2}   \left[\Delta(0)\right]^{\frac 32} \int d^du_1 \, d^du_2\, \left[3\,K\, \sqrt z\, +\, z^{\frac 32} \,{}_3F_2 \left(1,1,\frac 12;\frac 52,2;z\right)\right] \Delta(x_1-u) \,,
\end{align}
while for the other odd values of $n$
\begin{align}\label{G22noddB}
&\mathcal{G}^{(\epsilon^2,2)}_{n,\,B} =\frac{(-i)^n\,\epsilon^2 g^2 \mu^4}{2}\, \left[ \frac{2}{\Delta(0)}\right]^{\frac n2-2}\Gamma\left(\frac{n}{2}-1\right) \nonumber\\
&\times \int d^du_1 \, d^du_2\, \left[K\,\sqrt z\, - \left(\frac n3 -\frac{2}{3}\right) z^{\frac 32} \,{}_3F_2 \left(1,1,\frac n2;\frac 52,2;z\right) \right]\,\Delta(x_1-u_1) \prod_{h=2}^n \!\Delta(x_h-u_2)\,\, + {\rm perm.}\nonumber\\
&-\frac{(-i)^n\,\epsilon ^2 g^2 \mu ^4}{2}  \left[ \frac{2}{\Delta(0)}\right]^{\frac n2-2} \sum_{j=1}^{\frac{n-1}{2}} \Gamma (j) \Gamma \left(\frac{n}{2}-j-1\right)  \nonumber\\
&\times \int d^du_1 \, d^du_2\, \sqrt z\, {}_2F_1\left(j,\frac{n}{2}-j-1;\frac{3}{2};z\right)\, \prod_{i=1}^{2j+1}\Delta(x_i-u_1) \prod_{h=2j+2}^n \!\Delta(x_h-u_2)\,\, + {\rm perm.}\,\,,
\end{align}
and the final expressions for the odd Green's functions is obtained summing up  $\mathcal{G}^{(\epsilon^2,2)}_{n,\,A}$ and $\mathcal{G}^{(\epsilon^2,2)}_{n,\,B}$. Let us move now to the even Green's functions.  

\paragraph{Even Green's functions.}
Reading from (\ref{Lpk}) the expression for the terms $\mathcal{L}_{1\,0}\cdot\mathcal{L}_{1\,0}$ and  $\mathcal{L}_{1\,1}\cdot\mathcal{L}_{1\,1}$
we obtain:
\begin{align} \label{G22nEven}
\mathcal{G}_n^{(\epsilon^2,2)} (x_1,\dots,x_n) 
&= \frac{1}{Z_0} \int \mathcal{D}\phi\,\, e^{-S_0}\, \phi(x_1)\dots\phi(x_n)\, \times \nonumber\\
&\times \Bigg\{\frac{\epsilon^2}{32} g^2\mu^4 \int d^du_1d^du_2\, \phi^2(u_1)\, \log \left[\mu^{2-d}\phi^2(u_1)\right]\,  \phi^2(u_2)\, \log \left[\mu^{2-d}\phi^2(u_2)\right] \nonumber \\
&-\frac{\epsilon^2\pi^2}{32} g^2\mu^4 \int d^du_1d^du_2\, \phi(u_1) |\phi(u_1)|\, \phi(u_2) |\phi(u_2)| \Bigg\}\,.
\end{align}

We already encountered the first of the two terms in (\ref{G22nEven}) when we calculated the $\mathcal{G}_n$ at $\mathcal{O}(\epsilon^2)$ for the ordinary scalar theory $g\phi^2(\phi^2)^\epsilon$ in \cite{Branchina:2020jhd}.
%Following steps similar to those considered for the odd Green's functions calculated above, it turns out that 
The contribution from this term is given by the sum of all the connected diagrams built with a couple of $\mathcal{O}(\epsilon)$ effective vertices $\Pi_k^{(\epsilon)}$ with an even number of legs. Labelling with $A_1$ and $B_1$ the diagrams in which the number of external legs attached to each vertex is even and odd respectively, we have
\begin{align} \label{G22evennA1Pi}
\mathcal{G}^{(\epsilon^2,2)}_{n,\,A_1}&=\frac{1}{2} \sum_{j=0}^{n/2} \int d^du_1 \, d^du_2 \sum_{l=1}^{\infty} \,\Pi^{(\epsilon)}_{2j+2l} \,\, \Pi^{(\epsilon)}_{n-2j+2l}\, \frac{\Delta(u_1-u_2)^{2l}}{(2l)!}\nonumber\\
&\times \prod_{i=1}^{2j}\Delta(x_i-u_1) \prod_{h=2j+1}^n \!\Delta(x_h-u_2) \,\,\,\, + \, \binom{n}{2j} -1\,\, {\rm perm.}\,,
\end{align}
\begin{align} \label{G22evennB1Pi}
\mathcal{G}^{(\epsilon^2,2)}_{n,\,B_1} &=\frac{1}{2} \sum_{j=0}^{\frac{n}{2}-1} \int d^du_1 \, d^du_2\, \sum_{l=0}^{\infty} \,\Pi^{(\epsilon)}_{2j+2l+2} \, \Pi^{(\epsilon)}_{n-2j+2l} \,\frac{\Delta(u_1-u_2)^{2l+1}}{(2l+1)!}\nonumber\\
&\times \prod_{i=1}^{2j+1}\Delta(x_i-u_1) \prod_{h=2j+2}^n \!\Delta(x_h-u_2) \,\,\,\, + \, \binom{n}{2j\!+\!1} -1\,\, {\rm perm.}
\end{align}

We now replace  (\ref{Pi1even2}) and  (\ref{Pi1even}) for the effective vertices in  (\ref{G22evennA1Pi}) and (\ref{G22evennB1Pi}), and sum the series over $l$. Starting with the two-point Green's function we get
\begin{align}
\mathcal{G}^{(\epsilon^2,2)}_{2,\,A_1} &= \frac{\epsilon^2}{12} g^2\mu^4 \Delta(0) \int d^du_1 d^du_2\,\, \left[ 3\, K\, z \,-\, z^2 \, {}_3F_2\left(1,1,2;\, \frac 52, 3;\, z\right)\right]\,\Delta(x_1-u_1) \Delta(x_2-u_1) 
\end{align}
\begin{align}
\mathcal{G}^{(\epsilon^2,2)}_{2,\,B_1} &= \frac{\epsilon^2}{12} g^2\mu^4 \Delta(0) \int d^du_1 d^du_2\,\, \left[ 3\, K^2\sqrt{z} \,+\, 2\,z^{\frac 32} \, {}_3F_2\left(1,1,1;\, \frac 52, 2; \,z\right)\right] \,\Delta(x_1-u_1) \Delta(x_2-u_2)\,,
\end{align}
while for the even Green's functions with $n\geq4$ we have:

{\scriptsize
\begin{align}
	\mathcal{G}^{(\epsilon^2,2)}_{n,\,A_1} &= \frac{\epsilon^2}{6} g^2\mu^4 (-1)^{\frac{n}{2}+1} \left[\frac{2}{\Delta(0)}\right]^{\frac{n}{2}-2} \Gamma\left(\frac{n}{2}\right) \int d^du_1 d^du_2\,\, \left[ 3\,K \,z - \frac n2 z^2 \, {}_3F_2\left(1,1,\frac n2 +1;\, \frac 52 , 3;\, z\right)\right]\,\prod_{i=1}^{n} \Delta(x_i-u_2) \nonumber \\
	&+ \frac{\epsilon^2}{2} g^2\mu^4 (-1)^{\frac{n}{2}} \left[\frac{2}{\Delta(0)}\right]^{\frac{n}{2}-2} \Gamma\left(\frac{n}{2}-1\right) \int d^du_1 d^du_2\,\, z \, {}_3F_2\left[\left(1,1,\frac n2 -1\right); \left(\frac 32 , 2\right); z\right]\prod_{i=1}^{2} \Delta(x_i-u_1) \prod_{k=3}^{n} \Delta(x_k-u_2) + \text{perm.}\nonumber \\
	&+ \frac{\epsilon^2}{8} g^2\mu^4 (-1)^{\frac{n}{2}} \left[\frac{2}{\Delta(0)}\right]^{\frac{n}{2}-2} \sum_{j=2}^{\frac{n}{2}-2} \Gamma\left(j-1\right)\Gamma\left(\frac{n}{2}-j-1\right) \int d^du_1 d^du_2 \left[{}_2F_1\left(j-1,\frac n2 -j-1; \frac 12 ; z\right) - 1\right]\nonumber\\
	&\times\prod_{i=1}^{2j} \Delta(x_i-u_1) \prod_{k=2j+1}^{n} \Delta(x_k-u_2) + \text{perm.} 
\end{align}

\begin{align}
	\mathcal{G}^{(\epsilon^2,2)}_{n,\,B_1} &= \frac{\epsilon^2}{6} g^2\mu^4 (-1)^{\frac{n}{2}} \left[\frac{2}{\Delta(0)}\right]^{\frac{n}{2}-2} \Gamma\left(\frac{n}{2}-1\right) \int d^du_1 d^du_2\,\, \left[ 3\,K\,\sqrt{z} \,-\, \left(n-2\right) z^{\frac 3 2} \, {}_3F_2\left( 1,1,\frac n2;\, \frac 52, 2;\, z\right)\right] \,\nonumber\\
	&\times \Delta(x_1-u_1)\,\prod_{i=2}^{n} \Delta(x_i-u_2) + \text{perm.}\nonumber \\
	&+ \frac{\epsilon^2}{4} g^2\mu^4 (-1)^{\frac{n}{2}+1} \left[\frac{2}{\Delta(0)}\right]^{\frac{n}{2}-2} \sum_{j=1}^{\frac{n}{2}-2} \Gamma\left(j\right)\Gamma\left(\frac{n}{2}-j-1\right) \int d^du_1 d^du_2 \,\,\sqrt{z}\,\, {}_2F_1\left(j,\frac n2 -j-1; \frac 32 ; z\right) \nonumber\\
	&\times \prod_{i=1}^{2j+1} \Delta(x_i-u_1) \prod_{k=2j+2}^{n} \Delta(x_k-u_2) + \text{perm.} 
\end{align}
}

We have to consider now the second of the two terms in (\ref{G22nEven}). 
Using (\ref{mod}), we write this contribution to $\mathcal{G}^{(\epsilon^2,2)}_{n}$ as:
\begin{align}\label{pathodd}
&-\frac{\epsilon^2}{8} g^2\mu^4 \int d^du_1 d^du_2 \int_{0}^{\infty} dt_1 \sum_{r_1=0}^{\infty} \frac{(-1)^{r_1} t^{2r_1}}{(2r_1+1)!} \int_{0}^{\infty} dt_2 \sum_{r_2=0}^{\infty} \frac{(-1)^{r_2} t^{2r_2}}{(2r_2+1)!} \times\nonumber \\
&\times \frac{1}{Z_0} \int \mathcal{D} \phi\,\, e^{-S_0}\, \phi(x_1)\dots\phi(x_n)\, \phi^{2r_1+3}(u_1)\, \phi^{2r_2+3}(u_2)\,.
\end{align}

As before, we need first to calculate the connected component of the functional integral in (\ref{pathodd})  for each $r_1,\,r_2\geq 0$. There are two classes of diagrams, depending on whether the number of external legs attached to each vertex is even (let us call them $A_2$-type diagrams) or odd ($B_2$-type diagrams), and we will treat these two cases separately.

Starting with the $A_2$-type contribution, we have diagrams of the kind
\begin{equation} \label{diagG22nBE}
\begin{tikzpicture}[baseline=(m)]
\begin{feynman}[inline=(m)]
\vertex[dot] (m) at (-0.9, 0) {};
\vertex[label={center:{\footnotesize $u_1$}}] (m0) at (-0.9,-0.5) {};
\vertex[dot] (n) at (0.9, 0) {};
\vertex[label={center:{\footnotesize $u_2$}}] (n0) at (0.9,-0.5) {};
\vertex (a) at (-2.5,1) {$x_{1}$};
\vertex (b) at (-2.5,-1) {$x_{2j}$};
\vertex (c) at (2.5,1) {$x_{2j+1}$};
\vertex (d) at (2.3,-1) {$x_{n}$};
\vertex (l1) at (1.6,0.50) {};
\vertex (l2) at (1.6,-0.50) {};
\vertex (p1) at (-1.6,0.5) {};
\vertex (p2) at (-1.6,-0.5) {};
\vertex (q1) at (-1.2,0.65) {};
\vertex (q2) at (-0.6,0.65) {};
\vertex (t1) at (1.2,0.65) {};
\vertex (t2) at (0.6,0.65) {};
\diagram* {
	(a) -- (m) -- [out=70, in=50, loop, min distance=0.9cm] m -- [out=130, in=110, loop, min distance=0.9cm] m -- (b);
	(m) -- [half left,out=30, in=150, dashed](n);
	(m) -- [half left,out=-30, in=-150, edge label'={\scriptsize $2l+1$}, dashed](n);
	(m) -- (n);
	(c) -- (n) -- [out=70, in=50, loop, min distance=0.9cm] n -- [out=130, in=110, loop, min distance=0.9cm] n -- (d);	
	(l1) -- [thick,dotted,out=-60, in=60](l2);
	(p2) -- [thick,dotted,out=120, in=-120](p1);
	(q2) -- [thick,dotted,out=160, in=20, edge label'={\tiny $r_1\!+1\!-j\!-l$}](q1);
	(t2) -- [thick,dotted,out=20, in=160, edge label= {\tiny $r_2\!+1\!-\frac{n}{2}\!+j\!-l$}](t1);
};
\end{feynman}
\end{tikzpicture}
\end{equation}
where for any couple $r_1,\,r_2\geq0$, the only possible diagrams are those with values of $0\leq j\leq \frac{n}{2}$ and $l\geq0$,  satisfying the conditions
\begin{equation}\label{cond2}
\begin{cases}
r_1\geq j-1 +l  \\
r_2\geq \frac{n}{2}-j-1 + l
\end{cases}
\end{equation}
which imply that the number of internal lines is limited by
\begin{equation}\label{condmin}
l\leq Min[r_1+1-j,\,r_2+1-\frac{n}{2}+j]\equiv Min\,.
\end{equation}

Under these conditions, the sum of the diagrams of the type (\ref{diagG22nBE}) gives
\begin{equation}
\sum_{j=0}^{n/2}\sum_{l=0}^{Min} C_{njl}(r_1,r_2) \,\Delta(0)^{r+1-j-l}\, \Delta(0)^{r_2+1-\frac{n}{2}+j-l}\, \Delta(u_1-u_2)^{2l+1} \prod_{i=1}^{2j} \Delta(x_i-u_1) \prod_{k=2j+1}^{n} \Delta(x_k-u_2)\,+\,\text{perm.}
\end{equation}
where $C_{njl}(r_1,r_2)$ is the combinatorial factor coming from the contractions:
\begin{align}\label{combin2}
C_{njl}(r_1,r_2) &= \left[(2r_1+3)!! (r_1+1)_{j+l}\, 2^{j+l}\right]\left[(2r_2+3)!! (r_2+1)_{\frac{n}{2}-j+l}\, 2^{\frac{n}{2}-j+l}\right]\frac{1}{(2l+1)!} \nonumber\\
&= B_{2j+2l+1}(r_1)\,B_{n-2j+2l+1}(r_2) \,\frac{1}{(2l+1)!}\,.
\end{align}

In (\ref{combin2}) the terms  
$B_{2j+2l+1}(r_1)$ and $B_{n-2j+2l+1}(r_2)$ are the combinatorial coefficients (\ref{B}) found when we were dealing with $\mathcal{O}(\epsilon)$ one-vertex diagrams with an odd number of legs.
As for the odd Green's functions, this makes it possible to express the final result for the even  $\mathcal{G}_n$ in terms of diagrams with two effective-vertices $\Pi^{(\epsilon)}_k$.
Moreover, for the same reasons seen for the odd Green's functions, we can extend both the series over  $r_1$ and $r_2$   
starting from $r_1=r_2=0$, and the sum over $l$ up to infinity, regardless the conditions  (\ref{cond2}) and (\ref{condmin}).   
We can then write (in this case there is no need to consider $\mathcal G_2$ separately) 
\begin{align}
\mathcal{G}^{(\epsilon^2,2)}_{n,\,A_2} &= \frac{1}{2} \sum_{j=0}^{n/2} \int d^du_1 d^du_2 \sum_{l=0}^{\infty} \left[-i\frac{\epsilon}{2} g\mu^2 \int_{0}^{\infty} dt_1 \sum_{r_1=0}^{\infty} \frac{(-1)^{r_1} t^{2r_1}}{(2r_1+1)!}  B_{2j+2l+1}(r_1) \Delta(0)^{r_1+1-j-l} \right] \nonumber \\ 
&\times\left[-i\frac{\epsilon}{2} g\mu^2 \int_{0}^{\infty} dt_2 \sum_{r_2=0}^{\infty} \frac{(-1)^{r_2} t^{2r_2}}{(2r_2+1)!} B_{n-2j+2l+1}(r_2)\Delta(0)^{r_2+1-\frac{n}{2}+j-l}\right] \frac{\Delta(u_1-u_2)^{2l+1}}{(2l+1)!} \nonumber \\
&\times\prod_{i=1}^{2j} \Delta(x_i-u_1) \prod_{k=2j+1}^{n} \Delta(x_k-u_2) + \binom{n}{2j}-1\,\,\, \text{perm.}
\end{align}
where we easily recognize in the square brackets the $\mathcal{O}(\epsilon)$ effective vertices (\ref{Pi1odddef}) with an odd number of legs. Then 
\begin{align} \label{G22evennA2Pi}
\mathcal{G}^{(\epsilon^2,2)}_{n,\,A_2} &= \frac{1}{2} \sum_{j=0}^{n/2} \int d^du_1 d^du_2 \sum_{l=0}^{\infty} \Pi^{(\epsilon)}_{2j+2l+1} \Pi^{(\epsilon)}_{n-2j+2l+1} \frac{\Delta(u_1-u_2)^{2l+1}}{(2l+1)!} \nonumber \\
&\times \prod_{i=1}^{2j} \Delta(x_i-u_1) \prod_{k=2j+1}^{n} \Delta(x_k-u_2) + \binom{n}{2j}-1\,\, \text{perm.}
\end{align}
Inserting in (\ref{G22evennA2Pi}) the $\Pi^{(\epsilon)}_k$ as given in  (\ref{Pi1odd}), and performing the sum over $l$, we get 

\begin{align}\label{G22evennA2}
\mathcal{G}^{(\epsilon^2,2)}_{n,\,A_2} &= \frac{\epsilon^2}{4} g^2\mu^4 (-i)^{n+2} \left[\frac{2}{\Delta(0)}\right]^{\frac{n}{2}-2} \sum_{j=0}^{n/2} \Gamma\left(j-\frac{1}{2}\right) \Gamma\left(\frac{n}{2}-j-\frac{1}{2}\right) \nonumber \\
&\times \int d^du_1 d^du_2 \,\,\sqrt{z}\,\, {}_2F_1\left(j-\frac 1 2,\frac n 2 - j - \frac 1 2; \frac 3 2; z\right) \nonumber \\
&\times \prod_{i=1}^{2j} \Delta(x_i-u_1) \prod_{k=2j+1}^{n} \Delta(x_k-u_2) + \binom{n}{2j}-1\,\, \text{perm.}
\end{align}

Moving to the $B_2$ contribution, we now see that the diagrams are of the kind
\begin{equation}
\begin{tikzpicture}[baseline=(m)]
\begin{feynman}[inline=(m)]
\vertex[dot] (m) at (-0.9, 0) {};
\vertex[label={center:{\footnotesize $u_1$}}] (m0) at (-0.9,-0.5) {};
\vertex[dot] (n) at (0.9, 0) {};
\vertex[label={center:{\footnotesize $u_2$}}] (n0) at (0.9,-0.5) {};
\vertex (a) at (-2.5,1) {$x_{1}$};
\vertex (b) at (-2.5,-1) {$x_{2j+1}$};
\vertex (c) at (2.5,1) {$x_{2j+2}$};
\vertex (d) at (2.3,-1) {$x_{n}$};
\vertex (l1) at (1.6,0.50) {};
\vertex (l2) at (1.6,-0.50) {};
\vertex (p1) at (-1.6,0.5) {};
\vertex (p2) at (-1.6,-0.5) {};
\vertex (q1) at (-1.2,0.65) {};
\vertex (q2) at (-0.6,0.65) {};
\vertex (t1) at (1.2,0.65) {};
\vertex (t2) at (0.6,0.65) {};
\diagram* {
	(a) -- (m) -- [out=70, in=50, loop, min distance=0.9cm] m -- [out=130, in=110, loop, min distance=0.9cm] m -- (b);
	(m) -- [half left,out=30, in=150, dashed](n);
	(m) -- [half left,out=-30, in=-150, edge label'={\scriptsize $2l$}, dashed](n);
	(m) -- [out=20, in=160](n);
	(m) -- [out=-20, in=-160](n);
	(c) -- (n) -- [out=70, in=50, loop, min distance=0.9cm] n -- [out=130, in=110, loop, min distance=0.9cm] n -- (d);	
	(l1) -- [thick,dotted,out=-60, in=60](l2);
	(p2) -- [thick,dotted,out=120, in=-120](p1);
	(q2) -- [thick,dotted,out=160, in=20, edge label'={\tiny $r_1\!+1\!-j\!-l$}](q1);
	(t2) -- [thick,dotted,out=20, in=160, edge label= {\tiny $r_2\!+2\!-\frac{n}{2}\!+j\!-l$}](t1);
};
\end{feynman}
\end{tikzpicture}
\end{equation}

Following steps similar to those made for the $A_2$ contributions, we obtain as before an expression in terms of diagrams built with two $\mathcal{\mathcal{O}(\epsilon)}$ effective vertices $\Pi^{(\epsilon)}_k$ with an odd number of legs:
\begin{align} \label{G22evennB2Pi}
\mathcal{G}^{(\epsilon^2,2)}_{n,\,B_2} &= \frac{1}{2} \sum_{j=0}^{n/2-1} \int d^du_1 d^du_2 \sum_{l=1}^{\infty} \Pi^{(\epsilon)}_{2j+2l+1} \Pi^{(\epsilon)}_{n-2j+2l-1} \frac{\Delta(u_1-u_2)^{2l}}{(2l)!}\times \nonumber \\
&\times \prod_{i=1}^{2j+1} \Delta(x_i-u_1) \prod_{k=2j+2}^{n} \Delta(x_k-u_2) + \binom{n}{2j+1}-1\,\, \text{perm.}
\end{align}
Inserting as before the   $\Pi^{(\epsilon)}_k$ given in  (\ref{Pi1odd}), and performing the sum over $l$:
\begin{align}
\mathcal{G}^{(\epsilon^2,2)}_{n,\,B_2} &=  \frac{\epsilon^2}{8}g^2\mu^4 (-i)^{n} \left[\frac{2}{\Delta(0)}\right]^{\frac{n}{2}-2} \sum_{j=0}^{\frac n2-1} \Gamma\left(j-\frac{1}{2}\right) \Gamma\left(\frac{n}{2}-j-\frac{3}{2}\right) \times \nonumber \\
&\times \int d^du_1 d^du_2 \left[ {}_2F_1\left(j-\frac 1 2,\frac n 2 - j - \frac 3 2; \frac 1 2; \left(\frac{\Delta(u_1-u_2)}{\Delta(0)}\right)^2\right) -1 \right]\times \nonumber \\
&\times \prod_{i=1}^{2j+1} \Delta(x_i-u_1) \prod_{k=2j+2}^{n} \Delta(x_k-u_2) + \binom{n}{2j+1}-1\,\, \text{perm.}
\end{align}

The $\mathcal{O}(\epsilon^2)$ contribution to the even Green's functions from the two-vertex diagrams  is then obtained summing up the four terms $A_1,\,A_2,\,B_1$ and $B_2$.

\begin{comment}
%From .... the final result for the even Green's function is: %Collecting now all the contributions to $\mathcal{G}^{(\epsilon^2,2)}_{2}$ and $\mathcal{G}^{(\epsilon^2,2)}_{n}$ with generic even values $n$, we finally have:
\begin{align}
\mathcal{G}^{(\epsilon^2,2)}_{2}&=\mathcal{G}^{(\epsilon^2,2)}_{2,\,A_1}+\mathcal{G}^{(\epsilon^2,2)}_{2,\,A_2}+\mathcal{G}^{(\epsilon^2,2)}_{2,\,B_1}+\mathcal{G}^{(\epsilon^2,2)}_{2,\,B_2} \\
\mathcal{G}^{(\epsilon^2,2)}_{n}&=\mathcal{G}^{(\epsilon^2,2)}_{n,\,A_1}+\mathcal{G}^{(\epsilon^2,2)}_{n,\,A_2}+\mathcal{G}^{(\epsilon^2,2)}_{n,\,B_1}+\mathcal{G}^{(\epsilon^2,2)}_{n,\,B_2}\label{2vc}
\end{align}
\end{comment}

\paragraph{Analysis of the UV behaviour.}
The analysis of the UV behaviour of the two-vertex contributions to the Green's functions is easily performed if we refer to (\ref{G22oddnAPi}), (\ref{G22oddnBPi}), (\ref{G22evennA1Pi}), (\ref{G22evennB1Pi}), (\ref{G22evennA2Pi}) and (\ref{G22evennB2Pi}), where the $\mathcal{G}^{(\epsilon^2,2)}_{n}$ are expressed in terms of effective vertices $\Pi^{(\epsilon)}_k$ and loop integrals.
From the UV behaviour of the $\Pi^{(\epsilon)}_k$ (that is immediately read from (\ref{Pi1odd}) and (\ref{Pi1even})), and from the superficial degree of divergence of the loop integrals, we observe that in each of the series over $l$ the first term is the dominant one, so that the leading behaviour of the different contributions are:

\begin{comment}
From (\ref{G22oddnAPi}),(\ref{G22oddnBPi}),(\ref{G22evennA1Pi}),(\ref{G22evennB1Pi}),(\ref{G22evennA2Pi}) and (\ref{G22evennB2Pi}) we can now read the UV behaviour of the two-vertex contributions to the Green's functions.  
The analysis is easily performed by observing that the $\mathcal{G}^{(\epsilon^2,2)}_{n}$ are expressed in terms of effective vertices $\Pi^{(\epsilon)}_k$ and of loop integrals, and that in each series over $l$ the first term is the dominant one. 
The UV behaviour of the generic $\mathcal{G}^{(\epsilon^2,2)}_{n}$ is then straightforwardly obtained from the UV behaviour of the $\Pi^{(\epsilon)}_k$ (that we immediately read from (\ref{Pi1odd}) and (\ref{Pi1even})), and from the superficial degree of divergence of the loop integrals. For the different contributions we then obtain 	contenuto...
\end{comment}
%(in (\ref{G22oddnBUV}) we have to distinguish $n=2$ from $n\neq 2$)
\begin{align} \label{G22oddnUV}
\mathcal{G}^{(\epsilon^2,2)}_{n,\,A} &\sim D_{\Lambda}\,
\Delta(0)^{-\frac n2}  \log\Delta(0) \qquad \mathcal{G}^{(\epsilon^2,2)}_{n,\,B} \sim \Delta(0)^{1-\frac n2} \log\Delta(0)
\\
\label{G22evenn1UV}
\mathcal{G}^{(\epsilon^2,2)}_{n,\,A_1} &\sim D_{\Lambda}\,
\Delta(0)^{-\frac n2}  \log\Delta(0) \qquad \mathcal{G}^{(\epsilon^2,2)}_{n,\,B_1} \sim \begin{cases}
 \Delta(0)^{1-\frac n2} \log\Delta(0) \,\, &\text{for } n\neq 2\\
 \log^2\Delta(0) \,\, &\text{for } n=2
\end{cases} 
\\
\label{G22evenn2UV}
\mathcal{G}^{(\epsilon^2,2)}_{n,\,A_2} &\sim \begin{cases}
\Delta(0)^{1-\frac n2} \,\, &\text{for } n\neq 2\\
\log\Delta(0) \,\, &\text{for } n=2
\end{cases}
\qquad 
\mathcal{G}^{(\epsilon^2,2)}_{n,\,B_2} \sim D_{\Lambda}\, \Delta(0)^{-\frac n2} 
\end{align}
\begin{comment}
\begin{align} \label{G22oddnAUV}
\mathcal{G}^{(\epsilon^2,2)}_{n,\,A} &\sim \begin{cases}
\Lambda^{-(d-2)\frac n2} D_{\Lambda} \log\Lambda  \,\,\quad&{\rm for} \, d>2 \\
(\log \Lambda)^{-\frac n2} \log\left(\log\Lambda\right)  \,\,\quad&{\rm for} \, d=2\,
\end{cases}
\end{align}
\begin{align}\label{G22oddnBUV}
\mathcal{G}^{(\epsilon^2,2)}_{n,\,B} \sim \begin{cases}
\Lambda^{(d-2)\left(1-\frac n2\right)} \log\Lambda \\
(\log \Lambda)^{1-\frac n2} \log\left(\log\Lambda\right) 
\end{cases}
\qquad
\mathcal{G}^{(\epsilon^2,2)}_{2,\,B} \sim 
\begin{cases}
\log^2 \Lambda \,\,\,\,\,\,\,\,\, \quad \quad  \quad \quad{\rm for} \, d>2 \\
\log^2 \left(\log \Lambda\right) \,\quad\quad  \quad{\rm for} \, d=2
\end{cases}
\end{align}
\end{comment}
where $D_{\Lambda}$ is the superficial degree of divergence of the loop integral with two propagators:
\begin{align} \label{Dlambda}
D_{\Lambda}\sim\begin{cases}
{\rm const.} \quad&{\rm for} \,\, d\leq 3\\
\log\Lambda \quad&{\rm for} \,\, d=4\\
\Lambda^{d-4} \quad&{\rm for} \,\, d>4\,\,.\\
\end{cases}
\end{align}
\begin{comment}
Let us now study the UV behaviour of  $\mathcal{G}^{(\epsilon^2,2)}_{n}$with $n\geq 3$. This is easily done with the help of (\ref{G22evennA1Pi}), (\ref{G22evennB1Pi}), (\ref{G22evennA2Pi}) and (\ref{G22evennB2Pi}), where the different contributions to $\mathcal{G}^{(\epsilon^2,2)}_{n}$are given in terms of effective vertices $\Pi^{(\epsilon)}_k$ and loop integrals. We immediately see that  $\mathcal{G}^{(\epsilon^2,2)}_{n,\,A_1}$ and $\mathcal{G}^{(\epsilon^2,2)}_{n,\,B_1}$ have the same UV  behaviour of  
$\mathcal{G}^{(\epsilon^2,2)}_{n,\,A}$ and $\mathcal{G}^{(\epsilon^2,2)}_{n,\,B}\label{Gneps2}$, given in 
(\ref{G22oddnAUV}) and (\ref{G22oddnBUV}) respectively, while for $\mathcal{G}^{(\epsilon^2,2)}_{n,\,A_2}$ and $\mathcal{G}^{(\epsilon^2,2)}_{n,\,B_2}$ (in (\ref{G22evennAUV}) we have to distinguish $n=2$ from $n\neq 2$)

\begin{align}\label{G22evennAUV}
\mathcal{G}^{(\epsilon^2,2)}_{n,\,A_2} \sim \begin{cases}
\Lambda^{(d-2)\left(1-\frac n2\right)} \\
(\log \Lambda)^{1-\frac n2} 
\end{cases}
\qquad\mathcal{G}^{(\epsilon^2,2)}_{2,\,A_2} \sim \begin{cases}
\log \Lambda \,\,\,\,\,\,\,\,\, \quad \quad  \quad \quad{\rm for} \, d>2 \\
\log \left(\log \Lambda\right) \,\quad\quad  \quad{\rm for} \, d=2\,
\end{cases}
\end{align}
\begin{align}
\label{G22evennBUV}
\mathcal{G}^{(\epsilon^2,2)}_{n,\,B_2} &\sim \begin{cases}
\Lambda^{-(d-2)\frac n2} \,\, D_{\Lambda} \quad&{\rm for} \, d>2 \\
(\log \Lambda)^{-\frac n2} \quad&{\rm for} \, d=2\,,
\end{cases}
\end{align}
with $D_{\Lambda}$ given in (\ref{Dlambda}).
\end{comment}

Let us consider now the results
(\ref{G2vanish2}), (\ref{Gnvanish2}), (\ref{G22oddnUV}), (\ref{G22evenn1UV}) and (\ref{G22evenn2UV})
for the UV behaviour of all the one-vertex and two-vertex $\mathcal{O}(\epsilon^2)$ contributions to the Green's functions.
We see that $\mathcal{G}_1$ and $\mathcal{G}_2$ diverge. As it was the case at $\mathcal{O}(\epsilon)$, these divergences are trivially removed with the introduction of $\mathcal{O}(\epsilon^2)$ linear and quadratic counterterms, together with an additional $\mathcal{O}(\epsilon^2)$ wave function renormalization. This would be the starting point of a systematic renormalization of the theory at each order in $\epsilon$.

On the other hand, we also see that the $\mathcal{G}_n$ with $n\geq3$ vanish. Therefore,
putting these results together with those obtained at $\mathcal{O}(\epsilon)$, we conclude that up to $\mathcal O(\epsilon^2)$ {\it the theory is non-interacting}.
However, before drawing any conclusion on the original program of realizing a systematic renormalization of the theory within the framework of the logarithmic expansion, we have to 
consider the higher orders.

%Therefore, as long as we do not know the UV behaviour of the Green's function at the other orders in $\epsilon$, we cannot make any conclusion on the original program of realizing a systematic renormalization of the theory within the framework of the logarithmic expansion.  

%Therefore, we cannot make any sense of the would be renormalization of  $\mathcal{G}_1$ and $\mathcal{G}_2$.

\section{Higher order contributions to the $\mathcal G_n$} 
In this section we extend the analysis on the UV behaviour of the Green's functions to higher orders in $\epsilon$. 
When moving to a generic order $\epsilon^p$, we encounter  different contributions coming from diagrams having a number of vertices $V\leq p$. %This analysis is enormously simplified if once again we introduce effective vertices at each order in $\epsilon$.

First of all we note that, referring to Eq.\,(\ref{Lpk}), the terms that contribute to $\mathcal G_n^{(\epsilon^p)}$ are given by products of\,  $\mathcal{L}_{p_i\,k_i}[\phi(u_i)]$, with $i=1,\dots,V$. Each $\mathcal{L}_{p_i\,k_i}$ brings a factor either of the type $\phi^2 \log ^k \phi^2$ or of the type $\phi |\phi| \log ^k \phi^2$, that have to be treated with (\ref{replica}) and (\ref{mod}). When performing the contractions, the combinatorial factors always split in coefficients of the kind $B_n$ and $C_n$ in (\ref{B}) and (\ref{C}), encountered when considered one-vertex diagrams 
(see (\ref{combin}) and (\ref{combin2}) for the $\mathcal{O}(\epsilon^2)$ two-vertex contributions). 
This gives rise to a factorization that finally 
leads to express the $\mathcal{G}_n^{(\epsilon^p)}$ as a sum of all the possible diagrams built in terms of effective vertices. 
This general result is the great advancement brought by the introduction of the effective vertices.

As a consequence, the analysis of the UV behaviour of the Green's functions becomes an easy task even at higher orders in $\epsilon$. 
We only need to derive the UV behaviour of the generic $\mathcal{O}(\epsilon^p)$ effective vertex $\Pi^{(\epsilon^p)}_k$, and the superficial degree of divergence of the loop integrals due to the propagators that connect these vertices.

To this end, let us first calculate the generic $\mathcal{O}(\epsilon^p)$ effective vertices $\Pi_n^{(\epsilon^p)}$ by considering the one-vertex contribution to the $\mathcal{G}_n$ at order $\epsilon^p$. From (\ref{Lpk}) we have
\begin{align}\label{Gm1n}
&\mathcal G_{n}^{(\epsilon^p,1)}(x_1,\dots,x_n) 
= -\frac{1}{Z_0}\int \mathcal{D}\phi\, e^{-S_0}\phi(x_1)\dots\phi(x_n)\int d^du\, \mathcal{L}_{p} \nonumber\\
&= - \frac{g\mu^2}{2\, p!}\left(\frac{\epsilon}{2}\right)^p \int d^du \sum_{k=0}^{p} \binom{p}{k} \frac{1}{Z_0} \int\mathcal{D}\phi\, e^{-S_0} \phi(x_1)\dots\phi(x_n)\, \phi(u)^2 \log^{p-k}\left(\mu^{2-d}\phi(u)^2\right) \left(i\pi \frac{|\phi(u)|}{\phi(u)} \right)^{k}
\end{align}

Let us begin by considering even Green's functions, for which we only have even values of $k$, i.e. even powers of $\frac{|\phi|}{\phi}$. In this case:
\begin{align}\label{Gnparim}
&\mathcal G_{n}^{(\epsilon^p,1)}(x_1,\dots,x_n)=
-\frac 12 g \mu^2 \sum_{\substack{k=0\\k\,even}}^{p} \left(\frac{\epsilon}{2}\right)^p\frac{(i\pi)^{k}}{(p-k)!k!}
\nonumber\\
&\times \int d^du \frac{1}{Z_0}\int \mathcal{D}\phi\, e^{-S_0}\phi(x_1)\dots\phi(x_n) \phi(u)^2 \log^{p-k}\left(\mu^{2-d}\phi(u)^2\right)\,. 
\end{align}

By following similar steps to those employed in the previous sections, the path integral in (\ref{Gnparim}) is immediately calculated. We get
\begin{equation}\label{Gnovm}
\mathcal G_{n}^{(\epsilon^p,1)}(x_1,\dots,x_n)=\Pi_n^{(\epsilon^p)} \,\int d^du\, \prod_{i=1}^{n}\Delta(x_i-u)\,,
\end{equation}
that diagrammatically is written as
\begin{equation}\label{diagram1}
\begin{tikzpicture}[baseline=-3]
\begin{feynman}
\vertex[draw, circle, minimum size=0.5cm,very thick] (m) at (0, 0) {$\Pi_n^{(\epsilon^p)}$};
\vertex (a) at (-1.1,1.1) {\footnotesize$2$};
\vertex (b) at (-1.1,-1.1) {\footnotesize$1$};
\vertex (c) at (1.5,0) {\footnotesize$n$};
\vertex (l1) at (0.8,0.35) {};
\vertex (l2) at (-0.2,0.7) {};
\diagram* {
	(a) -- (m) -- (b);
	(m) -- (c);
	(l1) -- [thick,dotted,out=-280, in=-305](l2);
};
\end{feynman}
\end{tikzpicture} 
\end{equation}
with the effective vertex $\Pi_n^{(\epsilon^{p})}$ at $\mathcal{O}(\epsilon^p)$ given by 
\begin{align}\label{Pimevendef}
\Pi^{(\epsilon^p)}_n &\equiv - \frac{g\mu^2}{p!}\left(\frac{\epsilon}{2}\right)^p \left[\frac{2}{\Delta(0)}\right]^{\frac{n}{2}-1} \sum_{\substack{k=0\\k\,even}}^{p} \binom{p}{k} (i\pi)^k\,\lim_{N\to1} \frac{d^{(p-k)}}{dN^{(p-k)}} \left[ \left(2\mu^{2-d}\Delta(0)\right)^{N-1} \frac{\Gamma\left(N+\frac 12\right)\Gamma\left(N+1\right)}{\Gamma\left(\frac 32\right)\Gamma(N+1-\frac n2)}\right]
\end{align}

Concerning the odd Green's functions, in (\ref{Gm1n}) we only have odd values of $k$, i.e. odd powers of $\frac{|\phi|}{\phi}$, so that in this case

\begin{align}\label{onevertex2}
&\mathcal G_{n}^{(\epsilon^p,1)}(x_1,\dots,x_n)=
-\frac 12 g \mu^2 \sum_{\substack{k=0\\k\,odd}}^{p} \left(\frac{\epsilon}{2}\right)^p\frac{(i\pi)^{k}}{(p-k)!k!}
\nonumber\\
&\times \int d^du \frac{1}{Z_0}\int \mathcal{D}\phi\, e^{-S_0}\phi(x_1)\dots\phi(x_n) \phi(u)|\phi(u)| \log^{p-k}\left(\mu^{2-d}\phi(u)^2\right)\,.
\end{align}

Performing the path integral in (\ref{onevertex2}) by applying the same techniques developed in the previous sections, we find that the odd $\mathcal{G}_n^{(\epsilon^p,1)}$ are again given by (\ref{Gnovm}), with an effective vertex that has the same expression as in (\ref{Pimevendef}), with the difference that $k$ is odd rather than even.

From (\ref{Pimevendef}), and from the equivalent equation with $k$ odd, we find that the UV behaviour of the $\Pi_n^{(\epsilon^p)}$ is given by:
\begin{align}\label{PiUV}
\Pi_n^{(\epsilon^p)}\sim
\begin{cases}
\log^{\,p} \Delta(0) \qquad &{\rm for}\,n=2 \\\Delta(0)^{1-\frac n2} \log^{\,p-1} \Delta(0) \qquad &{\rm for}\,n\neq2
\end{cases}
\end{align}
where (as we already know) $\Delta(0)$ goes as $\log \Lambda$ for $d=2$, and as  $\Lambda^{d-2}$ for $d>2$. 
We note that, when  $p=1, 2$, Eq.\,(\ref{PiUV}) reduces to the cases previously studied cases. 

Let us consider now the UV behaviour of $\mathcal{O}(\epsilon^p)$ diagrams containing more than one effective vertex. 
As one-particle reducible diagrams contribute only to  dress legs, in the following we limit ourselves to consider only 1PI diagrams.

Indicating with $N_I$ the number of internal lines, the superficial degree of divergence $D$ of a generic loop integral is:
\begin{equation} \label{Loopscontr}
D=d(N_I-V+1)-2N_I\,,
\end{equation}
and the latter is (superficially) convergent when $D<0$.

Concerning the contribution of the effective vertices, we begin by noting that the total number of legs is $n+2N_I$.
Therefore the UV behaviour of the effective vertices is:
\begin{equation} \label{Picontr}
\prod_{i=1}^{V} \Pi^{(k_i)}_{m_i} \sim \prod_{i=1}^{V} \Delta(0)^{1-\frac{m_i}{2}} \log^{\,k_i-1} \Delta(0) \sim \Delta(0)^{V-\frac{n}{2}-N_i} 
\end{equation}
where in the last member of (\ref{Picontr}) we retain only the algebraic dependence on $\Delta(0)$. Actually we neglected a factor containing a power of $\log\Delta(0)$, as it is harmless for our analysis.
 
Let us distinguish the two cases $D<0$ and $D\geq0$.

\noindent (i) {\it Case $D<0$}. In this case the UV behaviour  dependence comes uniquely from the effective vertices, i.e. from (\ref{Picontr}). For any fixed value of $V$, the condition $D<0$ is verified when $V\leq N_I < \frac{d}{d-2}(V-1)$, where the lower limit comes from the fact that we are considering only 1PI diagrams.
The leading diagram (i.e. the one that is less suppressed in terms of $\Lambda$) is obtained when $N_i$ takes its minimal value, that is $N_I=V$. It goes as:
\begin{align}\label{caseD<0}
 \Delta(0)^{-\frac n2}
\end{align}

\noindent (ii) {\it Case $D\geq 0$}. This case does not exist in $d=2$ dimensions. For $d>2$, the UV behaviour of the diagram with $n$ external lines, $V$ vertices, and $N_I$ internal lines comes from both the effective vertices (\ref{Picontr}) and the loop integrals (see (\ref{Loopscontr})). It is:
\begin{equation}\label{vertexcutoff}
{\Lambda^{\frac{2-d}{2}n+d-2V}} \qquad {\rm for}\,d>2\,.
\end{equation}

From Eqs.\,(\ref{Gnovm}), (\ref{PiUV}),  (\ref{caseD<0}), and (\ref{vertexcutoff}) we see that at any order $\epsilon^p$ the dominant contibution to each of the Green's functions  $\mathcal{G}_n$ comes from the diagram with $V=1$, that is from the one-vertex diagram given by  (\ref{Gnovm}) and (\ref{PiUV}). In Sections 2 and 3 we have already seen that this is the case at order $\epsilon$ and $\epsilon^2$.

Moreover, from (\ref{Gnovm}) and (\ref{PiUV}) we see that the Green's functions $\mathcal G_1$ and $\mathcal G_2$ diverge at each order $\epsilon^p$. This is what we already found for the cases $p=1,2$.
These divergences are trivially removed with the introduction of linear and quadratic counterterms at each order $\epsilon^p$. 

Finally, (\ref{Gnovm}) and (\ref{PiUV}) also tell us that the Green's functions $\mathcal{G}_n$ with $n \geq 3$ vanish  for any dimension $d \geq 2$. Therefore: 
\begin{center}
{\it At any finite order in $\epsilon$, and for any $d \geq 2$, the theory is non-interacting}.
\end{center}

This is a disturbing, and at first sight surprising, result. However it is not difficult to understand its origin. 
%is worth noting that  Such a behaviour was already noted in \cite{Branchina:2020jhd} for the corresponding ordinary scalar theory, where it was found that
Actually, any finite order in $\epsilon$ gives nothing but an approximation to the interaction lagrangian that is truncated at a finite power of $\log(\phi)$. Due to the mild behaviour of the logarithm (and of its powers), we can easily understand that this is a ``too poor" truncation of the physical interaction, not sufficient to grasp enough of the quantum fluctuations, i.e. to guarantee the existence of non-trivial $S$-matrix elements.

It is worth to stress at this point that similar results were obtained by us for the corresponding $g \phi^2 (\phi^2)^{\epsilon}$ ordinary (hermitian) theory\cite{Branchina:2020jhd}. Performing the same expansion, we found that at each order in $\epsilon$ the theory $g \phi^2 (\phi^2)^{\epsilon}$ is non-interacting. 

This clearly shows that the fact  that the theory $g \phi^2 (i\phi)^{\epsilon}$ is
non-interacting at each order in $\epsilon$ has nothing to do with its non-hermitian nature. It is rather an intrinsic weakness of the logarithmic expansion. As just stressed, the reason is that at any finite order in $\epsilon$ this expansion
gives a ``too poor" truncation of the interaction. 

Moreover, the fact that at any order in $\epsilon$ the theory is non-interacting shows that the original program of a systematic renormalization of the theory at each order in $\epsilon$\cite{Bender:2018pbv} cannot be realized, actually looses its meaning. In particular, the renormalization of the field vacuum expectation value and of the mass, obtained through the introduction in the lagrangian of the linear and quadratic counterterms\, $\delta v\,\phi$ and  $ \frac12\delta m^2 \,\phi^2$ (considered above), has no physical significance. 

%In other words, as long as we stick on an expansion in powers of $\epsilon$, and consider finite orders in this expansion, there is no way to implement a systematic renormalization program for the theory. 
Physically speaking, there is no way of giving a meaning to the theory if we truncate it to a finite order in $\epsilon$.

We might still hope that a sensible definition could be given after resumming diagrams from each order in $\epsilon$. In this respect, it is clear that the first thing to do is to resum the leading contributions to the Green's functions, that as we have previously shown are the one-vertex diagrams (\ref{Gnovm}). This amounts at resumming the ladder of leading logarithms in (\ref{PiUV}). 

Before ending this section we observe that, in order to perform the resummations, it is useful, although not necessary, to go back to Eq.\,(\ref{Pimevendef}) (and the analogous one for odd values of $k$) and close the sum over $k$.  
To this end, defining 
\begin{align} \label{fnN}
f_n(N)= \left(2\mu^{2-d}\Delta(0)\right)^{N-1} \frac{\Gamma\left(N+\frac 12\right)\Gamma\left(N+1\right)}{\Gamma\left(\frac 32\right)\Gamma(N+1-\frac n2)} \times \begin{cases}
\cos(N\pi) \quad &\text{for $n$ even} \\
i\sin(N\pi) \quad &\text{for $n$ odd} 
\end{cases}
\end{align}
and noting that
\begin{equation}
\lim_{N \to 1}\frac{d^k}{dN^k} \sin(N\pi) = 
\begin{cases}
0 \\
i (i\pi)^k
\end{cases}
\qquad
\lim_{N \to 1}\frac{d^k}{dN^k} \cos(N\pi) = 
\begin{cases}
-(i\pi)^k \qquad&\text{for $k$ even}\\
0 \qquad&\text{for $k$ odd}
\end{cases}
\end{equation}
we can write the generic $n$-legs effective vertex at $\mathcal{O}(\epsilon^p)$ as
\begin{equation} \label{Pimn}
\Pi^{(\epsilon^p)}_n = g\mu^2 \left[\frac{2}{\Delta(0)}\right]^{\frac{n}{2}-1} \left(\frac{\epsilon}{2}\right)^p \frac{1}{p!} \lim_{N\to1} \frac{d^p}{dN^p} f_n(N)
\end{equation}

Let us proceed now to the resummations.

\section{Resummation of the one-vertex diagrams}

In the previous sections we have seen that, at any given order $\epsilon^p$, the leading contributions to the Green's functions $\mathcal G_n$ come from  the one-vertex diagrams (\ref{diagram1}).

Indicating their resummation  with $\mathcal G_{n}^{(1)}$: 
\begin{align} \label{Gresg}
\mathcal{G}_{n}^{(1)}(x_1,\dots,x_n)&=\sum_{p=1}^{\infty}\mathcal{G}_{n}^{(\epsilon^p,1)}(x_1,\dots,x_n)=\left(\sum_{p=1}^{\infty} \Pi_n^{(\epsilon^p)}\right)\int d^du\, \prod_{i=1}^{n} \Delta(x_i-u)\,. 
\end{align}

Eq.\,(\ref{Gresg}) shows that the resummation of the $\mathcal{G}_{n}^{(\epsilon^p,1)}$ actually amounts to resumming the $\Pi_n^{(\epsilon^p)}$, thus providing an $x$-independent (and then $p$-independent) approximation  $\Gamma^{(1)}_n$ to the $n$-points vertex function $\Gamma_n$.
From Eq.\,(\ref{Pimn}) we see that:
\begin{align}\label{Gamma1resum}
\Gamma^{(1)}_n \equiv \sum_{p=1}^{\infty} \Pi_n^{(\epsilon^p)} &= g\mu^2 \left[\frac{2}{\Delta(0)}\right]^{\frac{n}{2}-1} \sum_{p=1}^{\infty} \left(\frac{\epsilon}{2}\right)^p \frac{1}{p!} \left(\frac{d^p f_n(N)}{dN^p}\right)_{N=1}  \nonumber\\
&= g\mu^2 \left[\frac{2}{\Delta(0)}\right]^{\frac{n}{2}-1}\left[ f_n\left(1+\frac{\epsilon}{2}\right) - f_n(1)\right]
\end{align}
where the last step is possible as the function $f_n(N)$ is analytic on the positive real axis. From (\ref{fnN}), we finally get
\begin{align}\label{Gamma1resumfinal}
\Gamma^{(1)}_n = g\mu^2 \delta_{n,2}- g\mu^2 \left[\frac{2}{\Delta(0)}\right]^{\frac{n}{2}-1} \left(2\mu^{2-d}\Delta(0)\right)^{\frac{\epsilon}{2}} \frac{\Gamma\left(\frac{\epsilon+3}{2}\right)\Gamma\left(\frac{\epsilon+4}{2}\right)}{\Gamma\left(\frac 32\right)\Gamma\left(\frac{\epsilon+4-n}{2}\right)} \times \begin{cases}
\cos(\frac{\epsilon\pi}{2}) \quad &\text{for $n$ even} \\
i\sin(\frac{\epsilon\pi}{2}) \quad &\text{for $n$ odd} 
\end{cases}
\end{align}

Few comments are in order.
First of all we note that the resummation of the leading logarithms has drastically changed the UV behaviour of the Green's functions. In particular the dependence on powers of logarithms ($\log^p (\Delta(0))$ or $\log^{p-1} (\Delta(0))$, see Eq.\,(\ref{PiUV})) has been traded by the algebraic dependence $\Delta(0)^{\frac{\epsilon}{2}}$.
Interestingly, while at any finite order in $\epsilon$, the UV behaviour of $\mathcal{G}_n$ does not depend on $\epsilon$, after resummation it does.

Restricting ourselves to integer values of $\epsilon$, from  Eq.\,(\ref{Gamma1resumfinal}) we see that, due to the presence of the functions $\cos(\frac{\epsilon\pi}{2})$ and $\sin(\frac{\epsilon\pi}{2})$, the $\mathcal{G}_n$ for which $n$ has not the same parity of $\epsilon$ vanish. In addition, %among those $\mathcal{G}_n$ for which $n$ has the same parity of $\epsilon$, 
as the denominator contains the function  $\Gamma\left(\frac{\epsilon+4-n}{2}\right)$, only those $\mathcal G_n$ with $n<\epsilon+4$ are non-vanishing.

However, the condition $\epsilon < 2$ is necessary to restrict the field variable $\phi$ in the path integral that defines the Green's functions to real values, as required by the method followed in the present work.
Therefore, among the possible integer values of $\epsilon$, 
we can consider only  $\epsilon=1$. 
As it corresponds to the largely considered $\mathcal{PT}$-symmetric  $\frac12 ig\phi^3$ theory (see the lagrangian (\ref{eq:L})), this case is particularly relevant to study. 

Specifying then Eq.\,(\ref{Gamma1resumfinal}) to the $\epsilon=1$ case, and considering the amputated Green's functions, i.e. the vertex functions $\Gamma^{(1)}_n$, we see that the only non-vanishing  $\mathcal{G}_n$  are $\mathcal{G}_1$, $\mathcal{G}_2$, and  $\mathcal{G}_3$. In this case %, the $\Gamma_n^{(1)}$ are:
\begin{align}
\label{Gamma1g}
\mathcal{G}^{(1)}_1=\Gamma_1^{(1)} &=- \frac 3 2 i g\mu^{3-\frac d 2}  \Delta(0)= \,
\begin{tikzpicture}[baseline=-3]
\begin{feynman}
\vertex[dot] (m) at (0, 0) {};
\vertex (a) at (-0.8,0)  ;
\vertex (a1) at (-0.68,-0.12);
\vertex (a2) at (-0.52,0.12);
\vertex (b) at (-0.7,-0.7) ;
\diagram* {
	(a) -- (m);
	(m) -- [out=50, in=-50, loop, min distance=1.5cm] m;
};
\end{feynman}
\draw (a1) to (a2);
\end{tikzpicture}\\
\label{Gamma2g}
\mathcal{G}^{(1)}_2=\Gamma_2^{(1)} &= g \mu^2=\,
\begin{tikzpicture}[baseline=-3]
\begin{feynman}
\vertex[crossed dot, minimum size=0.2cm] (m) at (0, 0) {};
\vertex (a) at (-0.8,0)  ;
\vertex (a1) at (-0.68,-0.12);
\vertex (a2) at (-0.52,0.12);
\vertex (b) at (0.8,0)  ;
\vertex (b1) at (0.68,0.12);
\vertex (b2) at (0.52,-0.12);
\diagram* {
	(a) -- (m) -- (b);
};
\draw (a1) to (a2);
\draw (b1) to (b2);
\end{feynman}
\end{tikzpicture} 
\\
\label{Gamma3g}
\mathcal{G}^{(1)}_3=\Gamma_3^{(1)} &=- 3 i g \mu^{3-\frac d 2}= 
\begin{tikzpicture}[baseline=-3]
\begin{feynman}
\vertex[dot] (m) at (0, 0) {};
\vertex (a) at (0.8,0);
\vertex (a1) at (0.68,0.12);
\vertex (a2) at (0.52,-0.12);
\vertex (b) at (-0.4,-0.693) ;
\vertex (b1) at (-0.45,-0.4);
\vertex (b2) at (-0.18,-0.64);
\vertex (c) at (-0.4,0.693);
\vertex (c1) at (-0.45,0.4);
\vertex (c2) at (-0.18,0.64);
\diagram* {
	(a) -- (m) -- (b);
	(c) -- (m);
};
\end{feynman}
\draw (a1) to (a2);
\draw (b1) to (b2);
\draw (c1) to (c2);
\end{tikzpicture}\,.
\\
\label{Gamma4g}
\mathcal{G}^{(1)}_n=\Gamma_n^{(1)} &= 0 \,\,\, \,\,\,\,\,\,\,\,\,\,\,\,\,\,\,\,\,\,\,\,\,\, {\rm for}\,\, n\geq 4\,.
\end{align}

From Eqs.\,(\ref{Gamma1g})-(\ref{Gamma4g}) we see that for the theory $\frac 12 i g  \phi^3$ the resummation that we have just performed gives nothing but the trivial  $\mathcal{O}(g)$ results for the vertex functions $\Gamma_n$. This is why in the r.h.s. of these equations we have introduced ordinary weak-coupling diagrams (not to be confused with the diagrams of the logarithmic expansion considered up to now). 

More specifically: (i) $\Gamma_3^{(1)}$ is just the three-point vertex; (ii) $\Gamma_1^{(1)}$ is the usual shift in the vev generated by the $\mathcal{O}(g)$ tadpole diagram;
(iii) $\Gamma_2^{(1)}$ is the two-point vertex that simply comes from having inserted the $\mathcal{O}(\epsilon^0)$ of the interaction term in the free lagrangian (see Eq.\,(\ref {L0})), so generating the well known trivial reshuffling in the expansion. %, giving rise to $g\mu^2$ insertions in the progagator, as in (\ref{Gamma2g}).  

The resummation of the leading contributions to the Green's functions has then produced just trivial results, and this is quite deceptive. At the same time, these results show that, once we resort to resummations,  physically sensible (albeit up to now trivial) results can be obtained. Here we resummed diagrams with one effective vertex $\Pi^{(\epsilon^p)}_k$. In the next section we proceed to the resummation of diagrams with two effective vertices.

\section{Resummation of two-vertex diagrams} 

The contribution to the Green's function $\mathcal G_n$ at $\mathcal O(\epsilon^p)$ ($p\geq2$) from diagrams with two vertices comes from terms of the kind $\int d^du_1\,  \mathcal{L}_{k} \int d^du_2\,\mathcal{L}_{p-k}$. 
Indicating this term with $\mathcal{G}^{(\epsilon^p,2)}_{n}$
we have 
%the sum of all the the contributions of the kind $\int d^d u \mathcal L_p \int d^d w \mathcal L_k$. 
\begin{equation} \label{Gk2n}
\mathcal{G}^{(\epsilon^p,2)}_{n}(x_1,\dots,x_n)\,=\,\frac 12 \sum_{k=1}^{p-1} \frac{1}{Z_0} \int \mathcal D \phi\,e^{-S_0}\,\phi(x_1)\dots\phi(x_n) \int d^du_1\,d^du_2 \, \mathcal L_{k}[\phi(u_1)] \, \mathcal L_{p-k}[\phi(u_2)]
\end{equation}
\begin{comment}
\textcolor{red}{
We know that, when looking at one-vertex diagrams, each $\mathcal L_p$ gives rise to the effective vertices of the same order $\Pi^{(\epsilon^p)}_a$. Moreover, looking at the decomposition of the generic $\mathcal L_p$ in (\ref{Lpk}), we understand that the terms that are even (odd) in the field (i.e. those respectively of the kind $\phi^2 \log^q \phi^2$ and $\phi |\phi| \log^q\phi^2$) give rise to the effective vertices with an even (odd) number of legs. 
When looking at two-vertex diagrams, the factorization of the combinatorial factor explained in the previous sections implies that these contributions are organized as sums of all the diagrams built with two effective vertices coming from the Lagrangian terms considered.}

As there are different classes of diagrams whether we are dealing with even or odd Green's functions, let us treat the two cases separately.
\end{comment}
In order to perform the path integral in (\ref{Gk2n}), we need to distinguish the odd from the even Green's functions. Let us begin by considering the odd $\mathcal{G}_n$. 

\paragraph{Odd Green's functions.}
When the number of external legs is odd, the only possible two-vertex diagrams are those containing one effective vertex with an even number of legs and one with an odd number. Obviously each of them can come either from $\mathcal L_{k}$ or from $\mathcal L_{p-k}$. However, under the sum over $k$ these two cases bring the same contribution, thus giving a factor $2$ that cancels the factor $\frac{1}{2}$ in (\ref{Gk2n}).
Therefore (as done in section 3.2) we distinguish two classes of diagrams, namely those where the two effective vertices are connected with an even number of internal lines, and those connected by and odd number of lines, indicating them as $A$-type and $B$-type diagrams respectively. We have:

\begin{align} \label{Gk2oddnA}
&\mathcal{G}^{(\epsilon^p,2)}_{n,\,A} =  \sum_{k=1}^{p-1} \sum_{j=0}^{\frac{n-1}{2}} \sum_{l=1}^{\infty} \,\,\,\Pi^{(\epsilon^k)}_{2j+2l} \,\,\, \Pi^{(\epsilon^{p-k})}_{n-2j+2l}\,\,\frac{1}{(2l)!}\nonumber\\
&\times \left[\int d^du_1 \, d^du_2\, \prod_{i=1}^{2j}\Delta(x_i-u_1) \prod_{h=2j+1}^n \!\Delta(x_h-u_2) \,\, \Delta(u_1-u_2)^{2l}\,\, + \, \binom{n}{2j}-1 \, {\rm perm.}\right]
\end{align}
and 
\begin{align} \label{Gk2oddnB}
\mathcal{G}^{(\epsilon^p,2)}_{n,\,B} &= \sum_{k=1}^{p-1} \sum_{j=0}^{\frac{n-1}{2}} \sum_{l=0}^{\infty} \,\,\,\Pi^{(\epsilon^{k})}_{2j+2l+2} \,\,\, \Pi^{(\epsilon^{p-k})}_{n-2j+2l}\,\,\frac{1}{(2l+1)!}\nonumber\\
&\times \left[\int d^du_1 \, d^du_2\, \prod_{i=1}^{2j+1}\Delta(x_i-u_1) \prod_{h=2j+2}^n \!\Delta(x_h-u_2) \,\, \Delta(u_1-u_2)^{2l+1}\,\, + \,\, \binom{n}{2j\!+\!1}-1 \,\, {\rm perm.}\right]\,.
\end{align}
that correspond to diagrams of the kind:

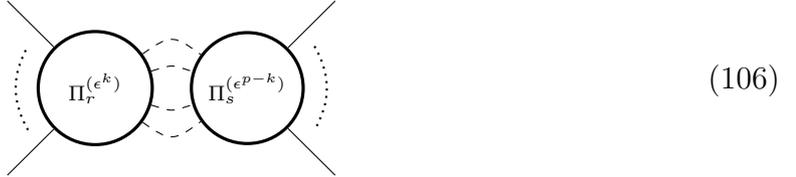
\begin{equation}
\begin{tikzpicture}[baseline=(m)]
\begin{feynman}[inline=(m)]
\vertex[draw, circle, minimum size=1.5cm,very thick] (m) at (-1, 0) {\scriptsize $\Pi_r^{(\epsilon^k)}$};
\vertex[draw, circle, minimum size=1cm,very thick] (n) at (1, 0) {\scriptsize $\Pi_s^{(\epsilon^{p-k})}$};	
\vertex (a) at (-2.3,1.3) {};
\vertex (b) at (-2.3,-1.3) {};
\vertex (c) at (2.3,1.3) {};
\vertex (d) at (2.3,-1.3) {};
\vertex (l1) at (1.8,0.7) {};
\vertex (l2) at (1.8,-0.7) {};
\vertex (p1) at (-1.8,0.7) {};
\vertex (p2) at (-1.8,-0.7) {};
\diagram* {
	(a) -- (m) -- (b);
	(m) -- [half left, out=17, in=163,dashed](n);
	(m) -- [half left, out=-17, in=-163,dashed](n);
	(m) -- [half left, out=36, in=144,dashed](n);
	(m) -- [half left, out=-36, in=-144,dashed](n);
	(c) -- (n) -- (d);	
	(l1) -- [thick,dotted,out=-60, in=60](l2);
	(p2) -- [thick,dotted,out=120, in=-120](p1);
};
\end{feynman}
\end{tikzpicture}
\end{equation}

Let us call $ \mathcal{G}^{(2)}_{n}\equiv \sum_{p=2}^{\infty} \mathcal{G}^{(\epsilon^p,2)}_{n}$
the sum of all these two-vertex diagrams, and
perform separately the resummation of $A$- and $B$- type diagrams. Let us begin with the former:
\begin{align} \label{Gng2e-1}
\mathcal{G}^{(2)}_{n,\,A} &=  \sum_{j=0}^{\frac{n-1}{2}} \sum_{l=1}^{\infty} \left\{\sum_{p=2}^{\infty}\,\sum_{k=1}^{p-1}\, \Pi^{(\epsilon^{k})}_{2j+2l} \, \Pi^{(\epsilon^{p-k})}_{n-2j+2l}\right\}\frac{1}{(2l)!} \left[I_{2l}(x_1{,\scriptstyle \dots},x_{2j};x_{2j+1}{,\scriptstyle \dots},x_n) \,\,+\,\, \binom{n}{2j}-1 \, {\rm perm.}\right]\,.
\end{align}
where we introduced the notation
\begin{equation}\label{intpropagator}
I_{r}(x_1,\dots,x_{2j};x_{2j+1},\dots,x_n)=\int d^du_1 \, d^du_2\, \prod_{i=1}^{2j}\Delta(x_i-u_1) \prod_{h=2j+1}^n \!\Delta(x_h-u_2) \,\, \Delta(u_1-u_2)^{r}\,.
\end{equation}
The double series in the curly brackets is nothing but the Cauchy product:
\begin{align} \label{splittingseries22}
\sum_{p=2}^{\infty}\,\sum_{k=1}^{p-1}\,\, \Pi^{(\epsilon^{k})}_{2j+2l} \,\, \Pi^{(\epsilon^{p-k})}_{n-2j+2l} \,=\, \left(\sum_{k=1}^{\infty}\Pi^{(\epsilon^{k})}_{2j+2l}\right) \left(\sum_{q=1}^{\infty}\Pi^{(\epsilon^{q})}_{n-2j+2l}\right)
\end{align}
where $q=p-k$, and we need that the two series in the r.h.s. are convergent, and at least one of them absolutely convergent. %From Eqs.\,(\ref{fnN}) and (\ref{Gamma1resum}) we see that the two series in the r.h.s. of (\ref{splittingseries22}) are absolutely convergent for $\epsilon<3$. 

We already summed in (\ref{Gamma1resumfinal}) the series in the round brackets.
They are Taylor expansions of functions analytic in the complex half-plane $Re(\epsilon)>-3$, due to the presence of $\Gamma\left(\frac{\epsilon+3}{2}\right)$ in (\ref{Gamma1resumfinal}), and then their radius of convergence in $\epsilon$ is $3$. Therefore the validity of (\ref{splittingseries22}) is guaranteed for $\epsilon<3$: in this case both series are absolutely convergent. 

For our scopes Eq.\,(\ref{splittingseries22}) is crucial, as it converts the resummation of diagrams with two effective vertices in the product of series whose terms are single effective vertices. Then, thanks to (\ref{Gamma1resum}), Eq.\,(\ref{splittingseries22}) results in the product of vertex functions $\Gamma^{(1)}_n$ (given in  (\ref{Gamma1resumfinal})).
From (\ref{Gng2e-1}) and (\ref{splittingseries22}) we then have:
\begin{equation} \label{Gng2e-2}
\mathcal{G}^{(2)}_{n,\,A}= \sum_{j=0}^{\frac{n-1}{2}} \sum_{l=1}^{\infty} \frac{\Gamma^{(1)}_{2j+2l}\,\,\Gamma^{(1)}_{n-2j+2l}}{(2l)!} \left[I_{2l}(x_1{,\scriptstyle \dots},x_{2j};x_{2j+1}{,\scriptstyle \dots},x_n) \,\,+\,\, \binom{n}{2j}-1 \, {\rm perm.}\right]\,.
\end{equation}
Following similar steps, the contribution from $B$-type diagrams is: 
\begin{align} \label{Gng2o-2}
\mathcal{G}^{(2)}_{n,\,B} =  \sum_{j=0}^{\frac{n-1}{2}} \sum_{l=0}^{\infty} \frac{\Gamma^{(1)}_{2j+2l+2}\,\, \Gamma^{(1)}_{n-2j+2l}}{(2l+1)!}\left[I_{2l+1}(x_1{,\scriptstyle \dots},x_{2j+1};x_{2j+2}{,\scriptstyle \dots},x_n) \,\,+\,\, \binom{n}{2j+1}-1 \, {\rm perm.}\right]\,.
\end{align}
and summing (\ref{Gng2e-2}) and (\ref{Gng2o-2}), we finally get the odd $\mathcal{G}^{(2)}_{n}$.

\paragraph{Even Green's functions.} 
When the number of external legs is even, there are two classes of diagrams: diagrams where the two effective vertices have an even number of legs (type $1$), and diagrams where both the effective vertices have an odd number of legs (type $2$).
Moreover, in both cases there are two possibilities: each of the  effective vertices can be connected either with an even ($A$-type diagrams) or with an  odd ($B$-type) number of external legs. Then we have 4 different classes of diagrams: $A_1$, $B_1$, $A_2$ and $B_2$ (the same classes that we encountered in section 3.2). 

Summing these contributions from all orders in $\epsilon$, and following similar steps to those made for the odd Green's functions, we find that all these contributions are easily expressed in terms of the resummed vertex functions $\Gamma^{(1)}_p$
\begin{align}\label{Gng2A1}
\mathcal{G}^{(2)}_{n,\,A_1}= \frac{1}{2}\sum_{j=0}^{\frac{n}{2}} \sum_{l=1}^{\infty} \frac{\Gamma^{(1)}_{2j+2l}\,\,\Gamma^{(1)}_{n-2j+2l}}{(2l)!} \left[I_{2l}(x_1{,\scriptstyle \dots},x_{2j};x_{2j+1}{,\scriptstyle \dots},x_n) \,\,+\,\, \binom{n}{2j}-1 \, {\rm perm.}\right]
\end{align}
\begin{align}
\mathcal{G}^{(2)}_{n,\,B_1}= \frac{1}{2} \sum_{j=0}^{\frac{n}{2}-1} \sum_{l=0}^{\infty} \frac{\Gamma^{(1)}_{2j+2l+2}\,\,\Gamma^{(1)}_{n-2j+2l}}{(2l+1)!} \left[I_{2l+1}(x_1{,\scriptstyle \dots},x_{2j+1};x_{2j+2}{,\scriptstyle \dots},x_n) \,\,+\,\, \binom{n}{2j+1}-1 \, {\rm perm.}\right]
\end{align}
\begin{align}
\mathcal{G}^{(2)}_{n,\,A_2}= \frac{1}{2}\sum_{j=0}^{\frac{n}{2}} \sum_{l=0}^{\infty} \frac{\Gamma^{(1)}_{2j+2l+1}\,\,\Gamma^{(1)}_{n-2j+2l+1}}{(2l+1)!} \left[I_{2l+1}(x_1{,\scriptstyle \dots},x_{2j};x_{2j+1}{,\scriptstyle \dots},x_n) \,\,+\,\, \binom{n}{2j}-1 \, {\rm perm.}\right]
\end{align}
\begin{align}\label{Gng2B2}
\mathcal{G}^{(2)}_{n,\,B_2}= \frac{1}{2} \sum_{j=0}^{\frac{n}{2}-1} \sum_{l=1}^{\infty} \frac{\Gamma^{(1)}_{2j+2l+1}\,\,\Gamma^{(1)}_{n-2j+2l-1}}{(2l)!} \left[I_{2l}(x_1{,\scriptstyle \dots},x_{2j+1};x_{2j+2}{,\scriptstyle \dots},x_n) \,\,+\,\, \binom{n}{2j+1}-1 \, {\rm perm.}\right]\,.
\end{align}
The sum of (\ref{Gng2A1})-(\ref{Gng2B2}) finally gives the even $\mathcal{G}^{(2)}_{n}$.

As in the previous section, we now specify to the interesting case of the $\frac12 i g \phi^3$ theory. We will see below that,  as it was the case for the resummation of diagrams with one effective vertex, again we get simple and trivial results of the weak-coupling expansion. 

Going to momentum space, and considering as usual amputated Green's functions, from each of the contributions in
(\ref{Gng2e-2})-(\ref{Gng2B2}) we get (the vertices with three and two legs of the weak-coupling expansion were already identified in the previous section):
%he following results (as in the previous section, the diagrams in the r.h.s of the equations below are weak-coupling diagrams): 

\begin{align}
&{\mathcal{G}}_{1,A}^{(2)} =\frac12 (-3 i g \mu^{3-\frac d 2})
(g\mu^2)\int \frac{d^dk}{(2\pi)^d}
\frac{1}{(k^2+M^2)^2} 
= 
\begin{tikzpicture}[baseline=-3]
\begin{feynman}
\vertex[dot] (m) at (0, 0) {};
\vertex[crossed dot, minimum size=0.2cm] (n) at (0.8, 0) {};
\vertex (a) at (-0.8,0);
\vertex (a1) at (-0.68,-0.12);
\vertex (a2) at (-0.52,0.12);
\diagram* {
	(a) -- (m);
	(m) -- [out=70, in=110] (n);
	(m) -- [out=-70, in=-110] (n);
};
\end{feynman}
\draw (a1) to (a2);
\end{tikzpicture}\label{weakphi3bis}
\\
&\mathcal{G}_{1,B}^{(2)} =\frac 12(g\mu^2)\frac{1}{M^2}
(-3 i g \mu^{3-\frac d 2})\int \frac{d^dk}{(2\pi)^d}
\frac{1}{k^2+M^2}
= \begin{tikzpicture}[baseline=-3]
\begin{feynman}
\vertex[dot] (m) at (0.2, 0) {};
\vertex[crossed dot, minimum size=0.20cm] (n) at (-0.2,0) {};
\vertex (a) at (-0.8,0) ;
\vertex (a1) at (-0.68,-0.12);
\vertex (a2) at (-0.52,0.12);
\vertex (b) at (-0.7,-0.7) ;
\diagram* {
	(a) -- (n)-- (m);
	(m) -- [out=50, in=-50, loop, min distance=1.4cm] m;
};
\end{feynman}
\draw (a1) to (a2);
\end{tikzpicture}\label{two}
\\
&\mathcal{G}_{3,B}^{(2)}(p_1,p_2,p_3) = (g\mu^2)\frac{1}{p_1^2+M^2}
(-3 i g \mu^{3-\frac d 2})\,+(p_1 \to p_2)+(p_1 \to p_3)\nonumber\\
&\qquad \qquad \,\quad \quad= 
\begin{tikzpicture}[baseline=-3]
\begin{feynman}
\vertex[dot] (m) at (0, 0) {};
\vertex[crossed dot, minimum size=0.20cm] (n) at (-0.45,0) {};
\vertex (a) at (-1,0);
\vertex (a1) at (-0.88,-0.12);
\vertex (a2) at (-0.72,0.12);
\vertex (b) at (0.4,-0.693) ;
\vertex (b1) at (0.45,-0.4);
\vertex (b2) at (0.18,-0.64);
\vertex (c) at (0.4,0.693);
\vertex (c1) at (0.45,0.4);
\vertex (c2) at (0.18,0.64);
\diagram* {
	(a) -- (n)--(m) -- (b);
	(c) -- (m);
};
\end{feynman}
\draw (a1) to (a2);
\draw (b1) to (b2);
\draw (c1) to (c2);
\end{tikzpicture}
\,+(p_1 \to p_2)+(p_1 \to p_3)
\label{three}
\\
&\mathcal{G}_{2,B_1}^{(2)} (p,-p) = (g\mu^2)\frac{1}{p^2+M^2}(g\mu^2)=
\begin{tikzpicture}[baseline=-3]
\begin{feynman}
\vertex[crossed dot, minimum size=0.2cm] (m) at (-0.3, 0) {};
\vertex[crossed dot, minimum size=0.2cm] (n) at (0.3, 0) {};
\vertex (a) at (-0.9,0)  ;
\vertex (a1) at (-0.78,-0.12);
\vertex (a2) at (-0.62,0.12);
\vertex (b) at (0.9,0)  ;
\vertex (b1) at (0.78,0.12);
\vertex (b2) at (0.62,-0.12);
\diagram* {
	(a) -- (m) -- (n) -- (b);
};
\draw (a1) to (a2);
\draw (b1) to (b2);
\end{feynman}
\end{tikzpicture}\label{four} 
\\
&\mathcal{G}_{2,A_2}^{(2)} =\frac 12 (-3 i g \mu^{3-\frac d 2})
\frac{1}{M^2}
(-3 i g \mu^{3-\frac d 2})
\int \frac{d^dk}{(2\pi)^d}
\frac{1}{k^2+M^2}
=
\begin{tikzpicture}[baseline=-3]
\begin{feynman}
\vertex[dot] (u) at (0, 0) {};
\vertex[dot] (w) at (0, 0.4) {};
\vertex (a) at (-0.8,0)  ;
\vertex (a1) at (-0.68,-0.12);
\vertex (a2) at (-0.52,0.12);
\vertex (b) at (0.8,0)  ;
\vertex (b1) at (0.68,0.12);
\vertex (b2) at (0.52,-0.12);
\diagram* {
	(a) -- (u) -- (b);
	(u) -- (w);
	(w) -- [loop, out=45, in=135, min distance= 0.9cm] w;
};
\draw (a1) to (a2);
\draw (b1) to (b2);
\end{feynman}
\end{tikzpicture}\label{five} 
\\
&\mathcal{G}_{2,B_2}^{(2)}(p,-p) =
\frac12 \,(-3 i g \mu^{3-\frac d 2})^2\,\int \frac{d^dk}{(2\pi)^d}
\frac{1}{k^2+M^2}\frac{1}{(k+p)^2+M^2}
= 
\begin{tikzpicture}[baseline=-3]
\begin{feynman}
\vertex[dot] (u) at (-0.4, 0) {};
\vertex[dot] (w) at (0.4, 0) {};
\vertex (a) at (-1,0)  ;
\vertex (a1) at (-0.88,-0.12);
\vertex (a2) at (-0.72,0.12);
\vertex (b) at (1,0)  ;
\vertex (b1) at (0.88,0.12);
\vertex (b2) at (0.72,-0.12);
\diagram* {
	(a) -- (u) --[out=50,in=130] (w) -- (b);
	(u) -- [out=-50,in=-130](w);
};
\draw (a1) to (a2);
\draw (b1) to (b2);
\end{feynman}
\end{tikzpicture} \label{six}
\\
&\mathcal{G}_{4,A_2}^{(2)}(s,t,u) =
\frac{(-3 i g \mu^{3-\frac d 2})^2}{s+M^2}\,+(s \to t)+(s \to u)\nonumber\\
& \qquad \qquad \quad \,=
\begin{tikzpicture}[baseline=-3]
\begin{feynman}
\vertex[dot] (u) at (-0.4, 0) {};
\vertex[dot] (w) at (0.4, 0) {};
\vertex (a) at (-1,0.6)  ;
\vertex (a1) at (-1.0,0.35);
\vertex (a2) at (-0.75,0.6);
\vertex (b) at (-1,-0.6) ;
\vertex (b1) at (-1.0,-0.35);
\vertex (b2) at (-0.75,-0.6);
\vertex (c) at (1,-0.6)  ;
\vertex (c1) at (1.0,-0.35);
\vertex (c2) at (0.75,-0.6);
\vertex (d) at (1,0.6)   ;
\vertex (d1) at (1.0,0.35);
\vertex (d2) at (0.75,0.6);
\diagram* {
	(a) -- (u) -- (b);
	(u) -- (w);
	(c) -- (w) -- (d);
};
\end{feynman}
\draw (a1) to (a2);
\draw (b1) to (b2);
\draw (c1) to (c2);
\draw (d1) to (d2);
\end{tikzpicture}
\,+(s \to t)+(s \to u)
\label{seven} \end{align}
where $s=(p_1+p_2)^2=(p_3+p_4)^2$, $t=(p_1-p_3)^2=(p_2-p_4)^2$ and $u=(p_1-p_4)^2=(p_2-p_3)^2$ with the conservation of the external momenta implied.

The above results are very deceptive.
Despite the additional effort needed to perform this second resummation, again we recover quite trivial weak-coupling results. 
The first four diagrams give corrections to the diagrams (\ref{Gamma1g})-(\ref{Gamma3g}) of the previous section, that are due to the insertion of $\Gamma_2^{(1)}$ in the propagators. The last three diagrams are typical contributions of the weak-coupling expansion at $\mathcal{O}(g^2)$. More specifically,  (\ref{five}) and (\ref{six}) are loop corrections to the propagator, while (\ref{seven}) is the tree level $2 \to 2$ particle scattering diagram.

\section{Summary and Conclusions}
In the present work we study the logarithmic expansion of the $\mathcal{PT}$-symmetric theory $g \phi^{2}(i\phi)^\epsilon$, applying techniques that we developed in a previous paper\cite{Branchina:2020jhd} to study the same expansion for an ordinary (hermitian) scalar theory .  %$g \phi^{2}(\phi^2)^\epsilon$.

The first order of this expansion %for the theory $g \phi^{2}(i\phi)^\epsilon$ 
was considered in\cite{Bender:2018pbv}, where it was suggested that such an expansion should be useful to implement a systematic renormalization of the theory at each order in $\epsilon$ (even though in that paper no attempt in this direction was made). 
This motivated our present work. 

Following our analysis\cite{Branchina:2020jhd}, we begin by introducing (at each order in $\epsilon$) effective vertices $\Pi_n$, with $n$ external legs,  that turn out to be powerful tools for our investigations.
The systematic analysis starts with the study of the $\mathcal O(\epsilon^2)$. We perform the calculation of {\it all the Green's functions} at this order, and find that we can write them in terms of effective vertices and loop integrals. More specifically, at this order we have two kind of contributions: (i) diagrams with one $\mathcal{O}(\epsilon^2)$ effective vertex; (ii) diagrams with two $\mathcal{O}(\epsilon)$ effective vertices linked by an infinite series of internal lines
(see Sections 3.1 and 3.2). 
We perform the resummation of these series and find a closed form for the $\mathcal{G}_n$ in terms of hypergeometric functions. 

The next step consists in studying the UV behavior of the Green's functions, and to this end we find it convenient to resort to the expansion in terms of effective vertices. The outcome of this analysis is that up to order $\epsilon^2$ the theory turns out to be non-interacting.

This pushed us to move a step further, extending the analysis of the UV behaviour of the Green's functions at all orders in $\epsilon$. To perform this analysis, it is extremely useful to write the different contributions to the $\mathcal{G}_n$ in terms of the two previously identified building blocks, effective vertices and loop integrals.
The outcome of this analysis is apparently very surprising: {\it at each finite order in $\epsilon$ the theory is non-interacting}.

We stress, however, that this result is surprising only in appearance. In fact, if we truncate the interaction to a finite power of $\log\phi$, that is what we do when considering a finite order in $\epsilon$, 
we actually implement a ``too poor" truncation of the physical interaction, not sufficient to guarantee the existence of non-trivial $S$-matrix elements. This is due to the mildness of the logarithm and of its powers.

The conclusion is that the original program (hope) of implementing the renormalization of the theory with the help of the systematic expansion in powers of $\epsilon$ looses completely its meaning. It cannot be realized. Thus, at face value, our conclusion appears to be rather negative. 

However, the UV analysis of the  Green's functions showed that, with increasing powers of $\epsilon$, the contributions to the $\mathcal{G}_n$ with $n\geq 3$ (that are the Green's functions relevant to establish whether the theory is interacting or not) become less and less suppressed. This suggested that we could resort to resummations to see whether this unpleasant result obtained for the finite orders could be overcome.

We then moved to consider the resummation of different contributions to the $\mathcal G_n$ at each order in $\epsilon$, starting with the leading ones. The latter coincide with  diagrams written in terms of a single effective vertex.  The result confirmed our expectation: after resummation some of the Green's functions with $n\geq 3$ no longer vanish. 

However, the result turned out to be quite trivial. In fact, resorting to the important $\mathcal{PT}$-symmetric $ig\phi^3$ theory, we see that, 
after the long detour implied by the method itself, 
%where we had to consider appropriate analytic extensions of functions appearing in the intermediate stages of the calculation, and to work hard to resum series, 
we simply get the trivial $\mathcal O(g)$ results of the weak-coupling expansion. 

Having considered the resummation of diagrams with a single effective vertex, we next moved to resum diagrams with two effective vertices, but again we got trivial results of the weak-coupling expansion. The outcome of these resummations has not really improved the status of the logarithmic expansion, as far as it concerns the possibility of using it to get non-trivial/non-perturbative results, as it was originally expected (hoped)\cite{Bender:1987dn,Bender:1988rq,Bender:2018pbv,Bender:1988ig}.
 
Naturally, there might still be the possibility that other resummations could provide less trivial results. 
%With the experience gained so far, however, we do not see how this could happen. To be more precise, we could imagine that resumming classes of diagrams different from those considered in the present work, we could get results that, once written in terms of the weak-coupling expansion, would result in resummations of diagrams with different powers of $g$, of the kind of Nambu-Jona-Lasinio, and/or Hartree-Fock resummations.
In any case, what we have definitely established with the present work is that the original idea to use the expansion in $\epsilon$ to implement a systematic renormalization of the theory cannot be implemented, actually it has no sense.

\paragraph*{Acknowledgments}
This work is carried out within the INFN project QFT-HEP and is supported in part by the Polish National Science Centre HARMONIA grant under contract UMO-2015/18/M/ST2/00518 (2016-2021).

\end{document}